\documentstyle[preprint,eqsecnum,aps,12pt]{revtex}

\author{A. P. de Almeida, F. T. Brandt and J. Frenkel\\
Instituto de F\'\i sica, Universidade de S\~ao Paulo,\\
S\~ao Paulo, 01498 SP, Brasil}
\title{Thermal matter and radiation in a gravitational field}

\begin{document}

\maketitle

\begin{abstract}
We study the one-loop contributions of matter and radiation to the
gravitational polarization tensor at finite temperatures. Using the
analytically continued imaginary-time formalism, the contribution of matter
is explicitly given to next-to-leading ($T^2$) order. We obtain an exact
form for the contribution of radiation fields, expressed in terms of
generalized Riemann zeta functions. A general expression is derived
for the physical polarization tensor, which is independent of the
parametrization of graviton fields. We investigate the effective
thermal masses associated with the normal modes of the corresponding
graviton self-energy.
\end{abstract}

\section{Introduction}

\label{intro}

Many properties of plasmas in thermal field theories can be understood from
the study of the polarization tensor evaluated at finite temperature \cite
{Weldon,KajantieKapusta,BraatenPisarski,FrenkelTaylor,EfratyNair}. This
tensor, which is the two-point correlation function, describes phenomena
such as the propagation of waves and dumping of fields in the plasma. In
thermal quantum gravity, the behavior of the polarization tensor is also of
interest, especially in connection with cosmological applications. If the
temperature $T$ is well below the Planck scale, perturbation theory can be
used to calculate the thermal Green functions. Thus, one obtains
loop-diagrams in which the internal lines represent matter and radiation in
thermal equilibrium, and the external lines represent the gravitational
fields. There has been a lot of work on hot quantum field theory in the
presence of a gravitational field \cite
{GrossPerryYaffe,GriboskyDonoghue,Rebhan,FrenkelTaylorBrandt}. Thus far
these investigations have been mainly restricted to the study of the
hard thermal loops contributions, which are obtained in the high
temperature limit.

The purpose of this work is to study the behavior of the graviton
polarization tensor at all temperatures, which might be useful in some
applications. Since these calculations are considerably more complicated
than those performed at high temperatures, we have restricted for
definiteness to work to one-loop order with thermal bosonic fields, which
may be of spin 0 or 1. The method we use is that of reference \cite
{FrenkelTaylorBrandt}, where the Green functions are related to a momentum
integral of the forward scattering amplitude of thermal particles in a
gravitational field. Then, the temperature-dependent part of the graviton
polarization tensor can be written at all temperatures in the form:
\begin{equation}
\label{PI1}\Pi ^{\mu \nu ,\alpha \beta }\left( k\right) =\frac 1{\left( 2\pi
\right) ^3}\int \frac{d^3q}{2\,Q}\frac 1{\exp \left( Q/T\right) -1}F^{\mu \nu
,\alpha \beta }\left( q,k\right) .
\end{equation}
Here $q_\mu \equiv \left( Q,\vec q\right) $ represents the on-shell momenta
of a thermal particle with mass $m$ and
energy $Q = \sqrt{\left| \vec q\right| ^2+m^2} $.
$F^{\mu \nu ,\alpha \beta }\left(q,k\right)$ is the forward
scattering amplitude, summed over the polarizations of thermal
particles, which is a covariant function of $q$ and the external momenta $k$%
. This temperature-independent amplitude is weighted in (\ref{PI1}) by the
Bose distribution factor. Because of the angular integrations, $\Pi ^{\mu \nu
,\alpha \beta }$ is no longer a Lorentz covariant function. It depends on
the time-like vector $u^\mu $, representing the local rest frame of the
plasma. For simplicity, we work in the comoving coordinate system where $%
u^\mu =\delta _0^\mu $. The above method simplifies very much the
calculations in the present case.

In Sec.\ref{sec2} we consider the contribution of matter particles
described by the scalar field $\phi $, coupled to a gravitational field.
The coupling characterized by the term $\xi R\phi ^2$ is included, where $%
\xi $ is a numerical factor and $R$ denotes the Ricci scalar. We verify that
$\Pi ^{\mu \nu ,\alpha \beta }$ satisfies the Ward identity which reflects
the invariance of the action under general coordinate transformations. We
obtain a general expression for the leading ($T^4$)
and next-to-leading ($T^2$)
contributions to the graviton polarization tensor. The special case when $%
\xi =-1/6$ and $m=0$ is of particular interest, since then the scalar action
is also invariant under conformal transformations \cite{BirrelDavies}. Due
to this invariance, $\Pi ^{\mu \nu ,\alpha \beta }$ satisfies in this case
a Weyl identity which is explicitly verified.

In Sec.\ref{sec3} we discuss the coupling of radiation fields which may
be photons or gluons, to a gravitational field. This coupling is also
invariant under general coordinate transformations as well as under
conformal transformations. We remark that the thermal contributions
associated with internal gauge fields represent gauge-invariant quantities.
The Ward and Weyl identities determine uniquely the ($T^4$) contributions,
which are the same for all thermal particles, apart from numerical factors
which count the number of degrees of freedom. Using general properties of
the forward scattering amplitude, we show that all other contributions can
be expressed in terms of just 2 parameters which are not fixed by the Ward
and Weyl identities. Rather, these parameters depend specifically on the
nature of thermal particles.

In Sec.\ref{sec4} we obtain a closed form expression for the
contributions of thermal radiation fields to the graviton polarization
tensor. We show that these can be expressed in terms of generalized Riemann
zeta functions $\zeta (-n,t)$ \cite{Gradshteyn} for natural values of
$n$, $t$ being a ratio of external momenta and the temperature. In the high
temperature limit, this expression yields a series of decreasing powers in the
temperature, which includes leading
$\left(T^4\right)$ and-next-to-leading $\left(T^2\right)$ contributions.
Some technical aspects which arise in the calculations are discussed in the
Appendices.

In Sec.\ref{sec5} we analyze the dependence on the parametrization of the
graviton fields, of the one-particle
irreducible (1PI) contributions to the graviton polarization tensor.
This behavior occurs generally
because of the non-vanishing of the thermal graviton 1-point function.
We show that the physical
polarization tensor, identified with the graviton
self-energy, is described by a traceless function which includes
contributions from thermal 1-point functions. A general expression for
the physical self-energy at finite temperature is derived, which is
independent of the graviton parametrization.

In Sec.\ref{sec6} we discuss the effective graviton propagator,
obtained by iterative insertions in the free propagator
of the physical self-energy. We
analyze, in the static limit, the corresponding poles which
describe three normal modes of dynamical screening. While one of the
modes remains unshielded, a non-vanishing screening mass $m_s^2=32 \pi G
\rho /3$ appears in the spatially transverse one,
where $\rho$ is the thermal energy
density. The spatially longitudinal
mode is characterized by an imaginary mass $m_J^2=-32 \pi G
\rho$, similar to the classical Jeans mass, indicating an instability
of thermal quantum gravity.

\section{Matter contributions to the polarization tensor}

\label{sec2}

We consider here thermal matter represented by scalar particles of mass m,
coupled to the gravitational field via the Lagrangian:
\begin{equation}
\label{MatterLagrangian}{\cal L}(x)=-\frac 12\sqrt{-g(x)}\left[ g^{\mu \nu}\,
\partial_\mu\phi \,\partial_\nu\phi - \left( m^2+\xi R\right) \phi ^2\right] ,
\end{equation}
and expand the metric tensor $g_{\mu \nu }$ in terms of the deviation from
the Minkowski metric $\eta _{\mu \nu }$:
\begin{equation}
\label{gdown}g_{\mu \nu }\equiv \eta _{\mu \nu }+\kappa \,h_{\mu \nu },
\end{equation}
where $\kappa =\sqrt{32\pi G}$. In order to derive the one-particle
irreducible (1PI) contributions to the thermal graviton 2-point function, we
consider the Feynman graphs shown in Fig.\ref{fig1}. According to Eq. (\ref
{PI1}), these can be expressed in terms of the forward scattering amplitude
of on-shell scalar particles, as indicated in Fig.\ref{fig2}. The
corresponding contributions to the amplitude can be expressed in terms of a
basis of 14 independents tensors $T_i^{\mu \nu ,\alpha \beta }(q,k)$, which
are symmetric under the interchanges $(\mu\leftrightarrow\nu)$,
$(\alpha\leftrightarrow\beta)$ and
$(\mu ,\nu )\leftrightarrow (\alpha,\beta )$.
These tensors are covariant functions of $q$ and $k$, being
polynomials of maximum degree 4 in the momenta. They can be obtained
from Table \ref{table1}, replacing the vectors $(X,Y)$ by the pair $(q,k)$.
With help of the Feynman rules given in Appendix \ref{appA}, it is
straightforward to obtain for the forward scattering amplitude the expression:
\begin{equation}
\label{MatterAmplitude}
\begin{array}{lll}
F^{\mu\nu,\,\alpha\beta}\left(q,k\right)=\displaystyle{\frac {\kappa^2}
{\left(k^2+2 k\cdot q\right)}}
&\left[\right.-
\left(\displaystyle{\frac 1 8}\xi \,{k^4} +
\displaystyle{\frac 1 4}\xi \,k^{2}\,k \cdot q\right)
T^{\mu\nu,\,\alpha\beta}_{1}-
\left(\displaystyle{\frac 1 4}\,k^{2} +
\displaystyle{\frac 1 2}\,k \cdot q\right)
T^{\mu\nu,\,\alpha\beta}_{2}+ \\
&
T^{\mu\nu,\,\alpha\beta}_{3}+
\left(\displaystyle{\frac 1 4}\,\xi \,{k^4} +
{{\xi }^2}\,{k^{4}} -
\displaystyle{\frac 1 2}\xi \,k^{2}\,k \cdot q +
\displaystyle{\frac 1 4}{{\left(k \cdot q\right)}^2}\right)
T^{\mu\nu,\,\alpha\beta}_{4}+ \\
& \left(\displaystyle{\frac 1 4}k^{2} +
\xi \,k^{2}\right)
T^{\mu\nu,\,\alpha\beta}_{5}+
\displaystyle{\frac 1 2}
T^{\mu\nu,\,\alpha\beta}_{7}+
\left(\displaystyle{\frac 1 8}\xi \,k^{2} +
\displaystyle{\frac 1 4}\xi \,k \cdot q\right)
T^{\mu\nu,\,\alpha\beta}_{8}- \\
& \xi
T^{\mu\nu,\,\alpha\beta}_{9}+
\displaystyle{\frac 1 4}
T^{\mu\nu,\,\alpha\beta}_{10}-
\displaystyle{\frac 1 2}\xi
T^{\mu\nu,\,\alpha\beta}_{11}+
{{\xi }^2}
T^{\mu\nu,\,\alpha\beta}_{12}- \\
& \left.\left(\displaystyle{\frac 1 4}\xi \,k^{2} +
{{\xi }^2}\,k^{2}\right)
T^{\mu\nu,\,\alpha\beta}_{13}+
\left(\displaystyle{\frac 1 2}\xi \,k^{2} -
\displaystyle{\frac 1 4}k \cdot q\right)
T^{\mu\nu,\,\alpha\beta}_{14}\right] \\
&+\left(k\leftrightarrow -k\right).
\end{array}
\end{equation}
To obtain the leading $(T^4)$ and the next to leading $(T^2)$ contributions
to the polarization tensor, we need to expand the energy $Q=\sqrt{\left|
\vec q\right| ^2+m^2}$ in powers of $(m^2/\left| \vec q\right| ^2)$, as well
as the Feynman denominators:
\begin{equation}
\label{ExpandDeno}\frac 1{k^2+2k\cdot q}=\frac 1{2k\cdot q}-\frac{k^2}{%
\left( 2k\cdot q\right) ^2}+\frac{k^4}{\left( 2k\cdot q\right) ^3}-\frac{k^6
}{\left( 2k\cdot q\right) ^4}+\cdots .
\end{equation}
The $T^4$ contributions come from terms in the forward amplitude (\ref
{MatterAmplitude}) which are homogeneous functions of $q$ of degree 2. These
are given by:
\begin{equation}
\label{LeadingMatter}
\begin{array}{l}
-T_2^{\mu \nu ,\,\alpha \beta }\left( q,k\right) -
\displaystyle\frac{k^2}{\left( k\cdot q\right) ^2}T_3^{\mu \nu ,\,\alpha
\beta }\left( q,k\right) +\displaystyle\frac 1{k\cdot q}T_7^{\mu \nu
,\,\alpha \beta }\left(k,q\right)
= \\ -\left( \eta ^{\nu \beta }q^\mu q^\alpha +\eta ^{\nu
\alpha }q^\mu q^\beta +\eta ^{\mu \beta }q^\nu q^\alpha +\eta ^{\mu \alpha
}q^\nu q^\beta \right) -
\displaystyle\frac{k^2}{\left( k\cdot q\right) ^2}q^\mu q^\nu q^\alpha
q^\beta  \\ \displaystyle\frac 1{k\cdot q}\left( q^\mu q^\nu q^\alpha
k^\beta +q^\mu q^\nu k^\alpha q^\beta +q^\mu k^\nu q^\alpha q^\beta +k^\mu
q^\nu q^\alpha q^\beta \right)
\end{array}
{}.
\end{equation}
Note that terms involving the parameter $\xi
$ do not contribute to (\ref{LeadingMatter}). These contribute only to
next-to-leading $(T^2)$ order, which result
from terms of degree zero in $q$ in
the forward amplitude. In order to find these contributions, we perform the
$\left|\vec q\right|$ integration in (\ref{PI1}) using the formulas:
\begin{equation}
\label{IntBose1}\int_0^\infty
\frac{\left| \vec q\right| d\left| \vec q\right| }{%
\left[ \exp \left( \left| \vec q\right| /T\right) -1\right] }=\frac{\pi ^2T^2%
}6,
\end{equation}
\begin{equation}
\label{IntBose2}\int_0^\infty
\frac{\left| \vec q\right| ^4d\left| \vec q\right| }{%
Q\left[ \exp \left( Q/T\right) -1\right] }=\frac{\pi ^4T^4}{15}-\frac{\pi
^2m^2T^2}4+\cdots .
\end{equation}
The angular integrals can be done using the methods described in \cite
{FrenkelTaylorBrandt}.
The result can be expressed in terms of the basis of 14
tensors $T^{\mu \nu ,\,\alpha \beta }_i\left( u,K\right) $, obtained from
Table \ref{table1}, where we replace the pair $\left( X,Y\right) $ by $%
\left( u,K\right) $. Here $u^\mu =\delta _0^\mu $ and
\begin{equation}
\label{BigK}K^\mu \equiv \frac{k^\mu }{\left| \vec k\right| }=\left( \frac{%
k_0}{\left| \vec k\right| },\,\hat k\right) \equiv \left( r,\,\hat k\right) .
\end{equation}
Then the 1PI contributions to the polarization tensor can be written up to
the next to leading order in the form:
\begin{equation}
\label{PI1T}\Pi ^{\mu \nu ,\,\alpha \beta }\left( k,m,\xi \right)
=\sum_{i=1}^{14}\Pi _i\left( r,K,\xi \right) T_i^{\mu \nu ,\,\alpha \beta
}\left( u,K\right) ,
\end{equation}
where:
\begin{equation}
\label{CoefPI1}\Pi _i\left( r,K,\xi \right) =\frac{\kappa ^2}{30}\left[ \pi
^2T^4\,l_i\left( r,K\right) +\left| \vec k\right| ^2T^2\,n_i\left(
r,K,\xi \right) +m^2T^2\,s_i\left( r,K,\xi \right) \right] +\cdots .
\end{equation}
The explicit form of the dimensionless functions $l_i(r,K)$, $n_i(r,K,\xi )$
and $s_i(r,K,\xi )$ are given in Appendix \ref{appB}. These exhibit, apart
from a logarithmic dependence in $r$, a polynomial behavior in $K$ of
maximum degree 10. The coefficients $l_i\left( r,K\right) $ which contribute
to the leading $(T^4)$ order, have been obtained previously \cite{Rebhan}
and are included here for completeness.

As a consequence of the invariance of the theory under general coordinate
transformations, the 1PI graviton 2-point function satisfies the Ward
identity:
\begin{equation}
\label{Ward}{\frac 2{\kappa}}
k_\nu \,\Pi ^{\mu \nu ,\,\alpha \beta }\left( k\right) =k^\mu
\,\Gamma ^{\alpha \beta }-k_\sigma \,\left( \Gamma ^{\alpha \sigma }\,\eta
^{\beta \mu }+\Gamma ^{\beta \sigma }\,\eta ^{\alpha \mu }\right) .
\end{equation}
Here $\Gamma ^{\alpha \beta }$ denotes the thermal graviton 1-point
function, which is given by:
\begin{equation}
\label{OnePoint}\Gamma ^{\alpha \beta }=\frac{\pi ^2T^4\kappa}
{180}\left( 4\,\delta
_0^\alpha \delta _0^\beta -\eta ^{\alpha \beta }\right) +\frac{m^2T^2\kappa}
{48}\left( \eta ^{\alpha \beta }-2\,\delta _0^\alpha \delta _0^\beta \right)
+\cdots .
\end{equation}
With the help of the expression given by Eqs. (\ref{PI1T})
and (\ref{CoefPI1}), the Ward
identity (\ref{Ward}) can be explicitly verified to this order. It is well
known \cite{BirrelDavies} that in the conformally coupled case, when $\xi
=-1/6$ and $m=0$, the action is also invariant under conformal
transformations given by:
\begin{equation}
\label{MetricTransf}\bar g_{\mu \nu }\left( x\right) =\Omega ^2\left(
x\right) \,g_{\mu \nu },
\end{equation}
\begin{equation}
\label{ScalarTransf}\bar \phi \left( x\right) =\Omega ^{-1}\left( x\right)
\,\phi \left( x\right) .
\end{equation}
In consequence of this invariance, the 1PI graviton 2-point function will
also satisfy the Weyl identity \cite{Rebhan,FrenkelTaylorBrandt}:
\begin{equation}
\label{Weyl}{\frac{1}{\kappa}}
\Pi _{\quad \;\,\,\,\sigma }^{\mu \nu ,\;\sigma }(k)=-\Gamma
^{\mu \nu }.
\end{equation}
This identity is explicitly verified by our expression for $\Pi ^{\mu \nu
,\,\alpha \beta }\left( k\right) $ $\left[ \text{Eq. }\left( \text{\ref{PI1T}%
}\right) \right] $ and for $\Gamma ^{\mu \nu }$ $\left[ \text{Eq. }\left(
\text{\ref{OnePoint}}\right) \right] $ evaluated at $\xi =-1/6$ with $m=0$.

\section{Radiation fields contributions to the polarization tensor}

\label{sec3}

In this section we analyze the contributions of spin 1 gauge fields, which
may be photons or gluons. Since for our purpose the self-interactions of the
Yang-Mills particles can be neglected, there is no loss of generality in
considering only the contribution of an Abelian field $A^{\mu \text{.}}$.
For non-Abelian fields the contributions are the same, up to an overall
color factor. The coupling of the gauge field $A^\mu $ is described by the
Lagrangian:
\begin{equation}
\label{RadiationLagrangian}{\cal L}_A=-\frac 14\sqrt{-g\left( x\right) }%
\,g^{\mu \nu }\,g^{\alpha \beta }\left( \partial _\mu A_\alpha -\partial
_\alpha A_\mu \right) \,\left( \partial _\nu A_\beta -\partial _\beta A_\nu
\right) .
\end{equation}
It is convenient for computational purposes to fix the gauge by choosing:
\begin{equation}
\label{GaugeFix}{\cal L}_{fix}=-{\frac {1}{2\alpha}} \sqrt{-g\left( x\right)
}\left( \nabla _\mu A^\mu \right) \left( \nabla_\nu A^\nu \right) ,
\end{equation}
where $\nabla _\mu $ is the covariant derivative. The corresponding Faddeev
Popov Lagrangian is given by:
\begin{equation}
\label{Ghost}{\cal L}_{FP}=g^{\mu \nu }\sqrt{-g\left( x\right) }\,\left(
\partial _\mu \bar \chi \right) \left( \partial_\nu \chi \right) ,
\end{equation}
where $\chi $ and $\bar \chi $ are the ghost fields. The form of the above
interactions is such that the theory is invariant under local coordinate
transformations, as well as under conformal transformations given by:
\begin{equation}
\label{ConformalTransf}\bar A_\mu \left( x\right) =A_\mu \left( x\right)
;\qquad \bar g_{\mu \nu }\left( x\right) =\Omega ^2\left( x\right) \,g_{\mu
\nu }.
\end{equation}
As we have seen, these invariances ensure the 1PI graviton 2-point function
to satisfy the Ward and Weyl identities given respectively by Eqs. (\ref
{Ward}) and (\ref{Weyl}). Here $\Gamma ^{\alpha \beta }$ is obtained
multiplying (\ref{OnePoint}) by a factor 2 and setting $m=0$.

With help of the Feynman rules listed in Appendix \ref{appA}, we can
evaluate the 1PI graphs contributing to $\Pi ^{\mu \nu ,\,\alpha \beta }$,
which are shown in Fig.\ref{fig3}. The diagrams contributing to the
corresponding forward amplitude are represented in Fig.\ref{fig4}. It is
important to note that the thermal contributions from internal gauge fields
represent gauge-independent quantities. We have verified this independence
explicitly, performing all computations in the general class of covariant
gauges defined by (\ref{GaugeFix}). The dependence on the gauge parameter
$\alpha$
cancels in the final expression of the forward scattering amplitude, which
is given by:
\begin{equation}
\label{RadiationAmplitude}
\begin{array}{lll}
F^{\mu \nu ,\,\alpha \beta }\left( q,k\right) = & \displaystyle\frac{2\kappa^2}
{k^2+2k\cdot q} & \left(
\displaystyle\frac 14\left( k\cdot q\right) ^2 T_1^{\mu \nu ,\,\alpha \beta}
\left(q,k\right)-
\displaystyle\frac{k\cdot q}2 T_2^{\mu \nu ,\,\alpha \beta}
\left(q,k\right)+
T_3^{\mu \nu,\,\alpha \beta }\left(q,k\right)\right. \\
&  & \left. -
\displaystyle\frac 14\left( k\cdot q\right) ^2 T_4^{\mu \nu ,\,\alpha \beta}
\left(q,k\right)-
\displaystyle\frac{k^2}4 T_5^{\mu \nu ,\,\alpha \beta}\left(q,k\right)+
\displaystyle\frac{k^4}8T_6^{\mu \nu ,\,\alpha \beta}
\left(q,k\right) \right. \\  &  & \left. +
\displaystyle\frac 12T_7^{\mu \nu,\,\alpha \beta }\left(q,k\right)+
\displaystyle\frac 12T_9^{\mu \nu ,\,\alpha \beta}\left(q,k\right)-
\displaystyle \frac{k^2}8 T_{14}^{\mu \nu ,\,\alpha \beta}
\left(q,k\right)\right) \\
& & + \left( k\leftrightarrow -k\right)
\end{array}
{}.
\end{equation}
At this point, it is interesting to compare (\ref{RadiationAmplitude}) with
the amplitude corresponding to the scalar case [Eq. (\ref{MatterAmplitude}%
)], evaluated for $m=0$ and $\xi =-1/6$. We see that in both amplitudes, the
coefficients of the tensors $T_i^{\mu \nu ,\,\alpha \beta }$ ($i=2,3,7$)
which contribute in the high temperature limit [cf. Eq. (\ref{LeadingMatter}%
)] are the same, up to a factor of 2 which counts the degrees of freedom of
a physical gauge particle. On the other hand, all other coefficients seen to
be different in general.

In order to understand this behavior we consider now the consequences of the
Ward (\ref{Ward}) and Weyl (\ref{Weyl}) identities on the structure of the
forward scattering amplitudes. To this end, we use the following
representation of the graviton one-point function:
\begin{equation}
\label{OnePointInt}\Gamma ^{\alpha \beta }={\frac {\kappa}{\left( 2\pi \right)
^3}}\int {\frac{d^3q}{Q}}
\frac 1{\exp \left( Q/T\right) -1}\,q^\alpha \,q^\beta .
\end{equation}
Then, we find from Eq. (\ref{PI1}) that the Ward and Weyl identities ensure
the forward scattering amplitudes to obey respectively the relations:
\begin{equation}
\label{WardScatering}{\frac {1}{\kappa^2}}
k_\nu F^{\mu \nu ,\,\alpha \beta }\left( q,k\right)
=k^\mu q^\alpha q^\beta -k\cdot q\left( q^\alpha \eta ^{\mu \beta }+q^\beta
\eta ^{\mu \alpha }\right) ,
\end{equation}
\begin{equation}
\label{WeylScatering}{\frac{1}{\kappa^2}}
F_{\;\quad \;\;\alpha }^{\mu \nu ,\,\alpha }\left(
q,k\right) =-2\,q^\mu q^\nu
\end{equation}
We will now investigate the constraints imposed by the relations (\ref
{WardScatering}) and (\ref{WeylScatering}) on the general form of the
amplitudes $F^{\mu \nu ,\,\alpha \beta }$. Since these are Lorentz covariant
functions of q and k, they can be expressed in terms of the tensor basis $%
T_i^{\mu \nu ,\,\alpha \beta }\left( q,k\right) $ as follows:
\begin{equation}
\label{StructureScatering}F^{\mu \nu ,\,\alpha \beta }\left( q,k\right)
=\kappa^2
\sum_{i=1}^{14}F_i\left( k^2,k\cdot q\right) T_i^{\mu \nu ,\,\alpha \beta
}\left( q,k\right) ,
\end{equation}
where $F_i$ are invariant functions of $k^2$ and $k\cdot q$. Inserting (\ref
{StructureScatering}) into the Ward identity (\ref{WardScatering}) and
identifying the coefficients of the independents tensor structures yields 10
relations among the $F_i$. Similarly the use of the Weyl identity (\ref
{WeylScatering}) gives 4 more relations. However, not all of these relations
are independent, so that we can express 11 functions $F_i$ in terms of the
remaining 3 as follows:
\begin{mathletters}
\label{Fs}
\begin{eqnarray}
\label{F1}F_1&=&{\frac {3k^4}{4}}F_{12}+k\cdot q F_{14} \\
\label{F2}F_2&=&-{\frac12}-{\frac {k^4} {2k\cdot q}}F_{11}+
{\frac {k^2} {k\cdot q}} F_{14} \\
\label{F3}F_3&=&-{\frac {k^2} {2(k\cdot q)^2}}-
{\frac {2 k^6} {(k\cdot q)^3}} F_{11}+
{\frac {k^4} {(k\cdot q)^3}} F_{14} \\
\label{F4}F_4&=&-{\frac {k^4}{2}} F_{12} - (k\cdot q) F_{14} \\
\label{F5}F_5&=&-{\frac {k^2} {(k\cdot q)}} F_{14} \\
\label{F6}F_6&=&{\frac {k^2} 2} F_{11}-F_{14}\\
\label{F7}F_7&=&{\frac {1} {2k\cdot q}}+
{\frac {2k^4} {(k\cdot q)^2}} F_{11}-
{\frac {k^2} {(k\cdot q)^2}} F_{14}  \\
\label{F8}F_8&=&-{\frac {k\cdot q} {2}} F_{11}-
{\frac {3\, k^2} {4} } F_{12} \\
\label{F9}F_9&=&-{\frac {2 k^2} {k\cdot q}} F_{11}+
{\frac {2} {k\cdot q}} F_{14} \\
\label{F10}F_{10}&=&-{\frac {3 k^2} {2 k\cdot q}} F_{11} \\
\label{F11}F_{13}&=&{\frac {k^2} {2}} F_{12}
\end{eqnarray}
\end{mathletters} We see that the invariance of the theory under local
coordinate and conformal transformations does not fix the functions $F_{11}$%
, $F_{12}$ and $F_{14}$. Further constraints are provided by the property of
the forward scattering amplitude of being a function with dimension of $(%
{\rm momenta})^2$, which is even under $\left( k\longleftrightarrow
-k\right) $. Furthermore, to one loop order
in perturbation theory this amplitude can have at
most one denominator involving $\left( k^2\pm 2k\cdot q\right) $. For
instance, these general properties require the functions $F_3$ and $F_7$ to
have the structure:
\begin{equation}
\label{F3c3}F_3=c_3\left( \frac 1{k^2+2k\cdot q}+\frac 1{k^2-2k\cdot
q}\right) ,
\end{equation}
\begin{equation}
\label{F7c7}F_7=c_7\left( \frac 1{k^2+2k\cdot q}-\frac 1{k^2-2k\cdot
q}\right) ,
\end{equation}
where $c_3$ and $c_7$ are constants. Furthermore, it follows that $F_2$ must
be an even function of $k$, having the structure:
\begin{equation}
\label{F2c2}F_2=c_2k\cdot q\left( \frac 1{k^2+2k\cdot q}-\frac 1{k^2-2k\cdot
q}\right) +c_2^{\prime }k^2\left( \frac 1{k^2+2k\cdot q}+\frac 1{k^2-2k\cdot
q}\right) ,
\end{equation}
where $c_2$ and $c_2^{\prime }$ are constants. Similar structures can be
found for all other functions appearing in Eqs. (\ref
{Fs}). These structures yield a set of
relations which must be satisfied identically in Eqs. (\ref
{Fs}), for all values of $q$ and $k$. In
this way, we find that the constants $c_2$, $c_3$ and $c_7$ are uniquely
determined as:
\begin{equation}
\label{c2c3c7}c_2=-\frac 12;\quad c_3=1;\quad c_7=\frac 12.
\end{equation}
Note that the functions $F_3$ and $F_7$, as well as the part of $F_2$ which
determine the $T^4$ contributions [cf. Eq.(\ref{LeadingMatter})] are now
uniquely fixed. This is in accordance with the argument \cite
{FrenkelTaylorBrandt} that all hard thermal particles should contribute the
same, up to a weight factor. The above relations imply further the equation:
\begin{equation}
\label{F14F11}F_{14}=2k^2F_{11}+\frac{k^2\left( k\cdot q\right) }{2\left[
k^4-4\left( k\cdot q\right) ^2\right] }.
\end{equation}
Using (\ref{F14F11}), we see from Eqs. (\ref
{Fs}) that the only independent functions
left over are $F_{11}$ and $F_{12}$. From the general properties of the forward
scattering amplitude, these functions must have the structure:
\begin{equation}
\label{F11c11}F_{11}=c_{11}\left( \frac 1{k^2+2k\cdot q}+\frac 1{k^2-2k\cdot
q}\right) ,
\end{equation}
\begin{equation}
\label{F12c12}F_{12}=c_{12}\left( \frac 1{k^2+2k\cdot q}-\frac 1{k^2-2k\cdot
q}\right) ,
\end{equation}
where $c_{11}$ and $c_{12}$ are constants which depend specifically on the
nature of the thermal particles. For instance, in the scalar case we get:
\begin{equation}
\label{Scalarc11c12}c_{11}=\frac 1{12};\qquad c_{12}=\frac 1{36},
\end{equation}
whereas in the case of internal gauge fields we find that:
\begin{equation}
\label{Gaugec11c12}c_{11}=c_{12}=0.
\end{equation}
The above relations explain the features of the forward scattering
amplitudes described by Eq. (\ref{MatterAmplitude}) [at $\xi =-1/6$ and $m=0$%
] and by Eq. (\ref{RadiationAmplitude}).

\section{Exact evaluation of radiation fields contributions}

\label{sec4}

We will now evaluate all finite-temperature contributions in closed
form, using the techniques described in the first paper of reference
\cite {AlmeidaFrenkelTaylor}. To this end, we express the 1PI graviton
2-point function in terms of the tensor basis $T_i^{\mu \nu ,\,\alpha
\beta }\left( u,K\right) $ in a way analogous to (\ref{PI1T}):
\begin{equation}
\label{PI2T}\Pi ^{\mu \nu ,\,\alpha \beta }\left( k\right)
=\sum_{i=1}^{14}\Pi _i\left( r,K\right) T_i^{\mu \nu ,\,\alpha \beta }\left(
u,K\right) .
\end{equation}

According to the discussion of the last section [cf.
Eq.(\ref{CoefPI1})], we can write the functions
$\Pi_i\left(r,K\right)$ as follows:
\begin{equation}
\label{CoefPI2}
\Pi _i\left( r,K\right) =\kappa ^2\left[\frac{\pi ^2T^4}{15}
l_i\left( r,K\right) +|\vec k|^2T^2N_i\left( r,K\right)\right] ,
\end{equation}
where the functions $l_i\left( r,K\right) $ are given in appendix
(\ref{appB}).  Our task is to determine the functions $N_i\left(
r,K\right) $, which should be non-leading in the high temperature
limit. For this, it is convenient to consider first the projections of
the graviton 2-point function into the tensor
basis $T_i^{\mu \nu,\,\alpha \beta }$:
\begin{equation}
\label{projections}\bar P_i\left( r,K\right) =\frac 18\Pi^{\mu \nu
,\,\alpha \beta }\left( k\right) T_{_i\,\mu \nu ,\,\alpha \beta }\left(
u,K\right) .
\end{equation}
Once we find these (see next), the functions $\Pi _i$ in (\ref{CoefPI2})
can be determined by the relation:
\begin{equation}
\label{SolCoefPI2}\Pi _i\left( r,K\right) =8(T_i^{\mu \nu ,\,\alpha \beta
}T_{_j\,\mu \nu ,\,\alpha \beta })^{-1}\bar P_j\left( r,K\right) \equiv
8\left( T_{ij}\right) ^{-1}\bar P_j\left(r,K\right),
\end{equation}
where $\left( T_{ij}\right) ^{-1}$ denotes the inverse of the matrix
$T_{ij}\equiv T_i^{\mu \nu ,\,\alpha \beta }T_{_j\,\mu \nu ,\,\alpha \beta }$.

We now proceed with the evaluation of the functions $\bar P_j\left(
r,K\right) $ in \ref{projections}. From Table \ref{table1}, we see
that for $j=4,5,\cdots ,14$ these involve the contraction of $\Pi
^{\mu \nu ,\,\alpha\beta }\left( k\right) $ with $\eta ^{\mu \nu }$,
$\eta ^{\alpha \beta }$ or with the external momenta. Using the Ward
(\ref{Ward}) and Weyl (\ref{Weyl}) identities, the corresponding
functions $\bar P_j\left( r,K\right) $ will be given by a linear
combination of graviton 1-point functions. These are proportional to
$T^4$ [cf. Eq. (\ref{OnePoint}) with $m=0$], and so will contribute
only to the functions $l_i\left( r,K\right) $ in (\ref{CoefPI2}). The
functions
$N_i\left( r,K\right) $ are determined from the contributions
corresponding to $\bar P_j\left( r,K\right) $ ($j=1,2,3$), which are
not proportional to $T^4$. These contributions, which we denote by
$P_j$ ($j=1,2,3$) can be found from Eq. (\ref{projections}) by using
for the graviton 2-point function the expression (\ref{PI1}).
Substituting here the expression (\ref{RadiationAmplitude}) for the
forward scattering amplitude, we are lead to integrals of the form:
\begin{equation}
\label{IntegralsIS}I_S\left( k,T\right) =\frac 1{\left( 2\pi \right) ^3}\int
\frac{d^3q}{2Q}\frac{Q^S}{\exp \left( Q/T\right) -1}
\left( \frac 1{k^2+2k\cdot q}+
       \frac 1{k^2-2k\cdot q}\right) ,
\end{equation}
where $S=0,2,4$ and $Q=\left|\vec q\right|$.
In terms of $x=\cos\left(\theta\right)$,
where $\theta $ is the angle between $\vec k$ and $\vec q$, we find
that the above expression becomes:
\begin{equation}
\label{IntegralsISAngular}I_S\left( k,T\right) =-\frac 1{\left( 2\pi \right)
^2}\frac{k^2}4\int_{-1}^{\,\,1}\frac{dx}{\left( k_0-\left| \vec
k\right| x\right) ^2}\int_0^\infty dQ\frac{Q^{S+1}}{\exp \left(
Q/T\right) -1}\frac 1{Q^2+\left( 2\pi Ty\right) ^2},
\end{equation}
where,
\begin{equation}
\label{yDefinition}y\equiv \frac 1{4i\pi T}\frac{k^2}{k_0-\left| \vec
k\right| x}.
\end{equation}
Apart from simple functions, the integration in
(\ref{IntegralsISAngular}) can be reduced to the basic integral
\cite{Gradshteyn}:
\begin{equation}
\label{BasicIntegral}I\left( y\right) =\int_0^\infty \frac{QdQ}{Q^2+\left(
2\pi Ty\right) ^2}\frac 1{\exp \left( Q/T\right) -1}=
\frac12\Theta[{\rm Re}(y)]\left(\ln y-\frac 1{2y}-\psi\left(y\right)
\right)+(y\leftrightarrow-y),
\end{equation}
where $\psi(y)=\displaystyle\frac d{dy}\ln\Gamma\left(y\right)$
denotes the Euler psi function. The real-time limit of the Green's
function can be obtained from the analytically continued
imaginary-time formalism via the prescription\cite{LandsmanVeert}
$k_0=(1+i\varepsilon)K_0$, where $\varepsilon\rightarrow0^+$ and $K_0$
is real. With the presence of the $i\varepsilon$ factor being
understood, we find in (\ref{BasicIntegral}) that ${\rm
Re}(y)=\varepsilon'{\rm Re}(k_0)$, with $\varepsilon'\rightarrow0^+$.

Many of the angular integrations in (\ref{IntegralsISAngular}) can be
easily done in terms of elementary functions, after changing variables
from $x$ to $y$. The most difficult one involve an integrand
containing $\psi\left(y\right)$ multiplied by a power of $y^n$, for
$n=0,1,2,3,4$. The relevant integrals can be put in the form:
\begin{equation}
\label{DefJ}
J_n\equiv\Theta[{\rm Re}(k_0)]\left[J_n(t(k_0))-J_n(-t(-k_0))\right]+
\left[k_0\leftrightarrow-k_0\right],
\end{equation}
where
\begin{equation}
\label{JnIntegrals}
J_n(t)=\left( \frac {i\pi T}{|\vec k|}\right) ^{n-1}\int_C^ty^n
\left[\ln(y)-\frac1{2y}-\psi(y)\right]dy.
\end{equation}

The choice of $C$ is immaterial, since any constant is irrelevant for
our purposes because it cancels out in the expression (\ref{DefJ}).
Here $t\left(k_0\right)$ and $-t\left(-k_0\right)$
denote the limits of the $y$-integration
corresponding respectively
to $x=1$ and $x=-1$ in Eq. (\ref{yDefinition}). Hence:
\begin{equation}
\label{tDefinition}t\left( k_0\right) =\frac{k_0+\left| \vec
k\right| }{4i\pi T}.
\end{equation}

These integrals can be expressed in closed form in terms of derivatives of
generalized Riemann zeta functions $\zeta \left( -n,t\right) $ for
natural values of $n=0,1,2,3,4$. For instance we have that:
\begin{equation}
\label{ExpampleOfJ}
J_0\left(t\right)=\frac{|\vec k|}{i\pi T}\int^t
\left[\ln(y)-\frac1{2y}-\psi(y)\right]dy=
\frac{|\vec k|}{i\pi T}\left[t\ln t-\zeta^{\prime}\left(0,t\right)-t
-\frac12\ln t\right],
\end{equation}
\begin{equation}
J_1(t)=\int^t y\left[\ln(y)-\frac1{2y}-\psi(y)\right]dy=
t\zeta^\prime(0,t)-\zeta^\prime(-1,t)+\frac{t^2}2\ln t-\frac34t^2.
\end{equation}

In the above expressions, the derivative is taken with respect to the
first argument of the generalized zeta function.  The functions
$J_n(t)$ are discussed in more generality in Appendix C [cf. Eq.
(\ref{eqj})].  In this way, we find that the functions $P_j$
($j=1,2,3$) in Eq.  (\ref{projections}) are related to $J_n$ in
(\ref{DefJ}) as follows:
\begin{equation}
\label{p1}P_1\left( r,K\right) =-\frac{K^2}{96}-\frac{K^4}{32}J_0,
\end{equation}
\begin{equation}
\label{p2}P_2=\frac{K^2}{48}L\left( r\right) -\frac{K^4}{32}J_0
-r\frac{K^2}4J_1-\frac{K^2}2J_2,
\end{equation}
\begin{equation}
\label{p3}P_3=-\frac{1}{576}+\frac{K^2}{192}L\left( r\right)
-\frac{K^4 }{256}J_0-r\frac{K^2}{16}J_1-\frac{2+3K^2}8J_2-rJ_3-J_4,
\end{equation}
where we have defined:
\begin{equation}
\label{LDefinition}
L\left( r\right) =\frac r2\ln \frac{r+1}{r-1}-1.
\end{equation}
With help of these relations, the functions $N_i\left( r,K\right) $ can be
explicitly determined from Eqs. (\ref{CoefPI2}) and (\ref{SolCoefPI2}).
After a straightforward calculation we obtain that:
\begin{mathletters}
\begin{eqnarray}
\label{N1}N_1(r,K)&=&P_1+K^2P_2+K^4P_3 \\
\label{N2}N_2(r,K)&=&K^2P_1+2K^4P_2+5K^6P_3 \\
\label{N3}N_3(r,K)&=&K^4P_1+5K^6P_2+35K^8P_3 \\
\label{N4}N_4(r,K)&=&-P_1-K^2P_2+K^4P_3 \\
\label{N5}N_5(r,K)&=&-K^2P_1-K^4P_2+5K^6P_3 \\
\label{N6}N_6(r,K)&=&-r\left(P_1+2K^2P_2+5K^4P_3\right)\\
\label{N7}N_7(r,K)&=&-r\left(K^2P_1+5K^4P_2+35K^6P_3\right)\\
\label{N8}N_8(r,K)&=&P_1+\left(1+2K^2\right)P_2+\left(4K^2+5K^4\right)P_3 \\
\label{N9}N_9(r,K)&=&\left(2+K^2\right)P_1+\left(6K^2+5K^4\right)P_2+
\left(30K^4+35K^6\right)P_3\\
\label{N10}N_{10}(r,K)&=&K^2P_1+\left(3K^2+5K^4\right)P_2+\left(30K^4+35K^6
\right)P_3\\
\label{N11}N_{11}(r,K)&=&-r\left[P_1+\left(2+5K^2\right)P_2+
\left(20K^2+35K^4\right)P_3\right] \\
\label{N12}N_{12}(r,K)&=&P_1+\left(4+5K^2\right)P_2+\left(8+40K^2+
35K^4\right)P_3\\
\label{N13}N_{13}(r,K)&=&-P_1-K^2P_2+\left(4K^2+5K^4\right)P_3\\
\label{N14}N_{14}(r,K)&=&r\left(P_1+K^2P_2+5K^4P_3\right)
\end{eqnarray}
\end{mathletters}

At high temperatures, the Riemann zeta functions can be expanded in a
power series in $t$, as shown in Eq. (\ref{expzl}). Then, in the high
temperature domain, we can express the functions $P_j$ ($j=1,2,3$) as
a series of powers of $(1/T)$. The dominant terms
in these series are given by:
\begin{eqnarray}
\label{p1t2}
P\,_{1}&=&-\frac{K^2}{96}\\
\label{p2t2}
P\,_{2}&=&\frac{K^2}{48}L(r)\\
\label{p3t2}
P\,_{3}&=&-\frac{1 - 3K^2L(r)}{576}
\end{eqnarray}

Although the above contributions are gauge invariant, they do not
directly describe the physical properties of the plasma at finite
temperatures in a gravitational field. This problem is related to
the fact that the thermal graviton $2$-point function depends on the
choice of the basic graviton fields.

\section{The graviton self-energy at finite temperature}

\label{sec5}

In thermal quantum field theory the 1PI contributions to the graviton
2-point function are in general dependent on the parametrization of the
graviton fields. However, as shown in the second work of reference \cite
{AlmeidaFrenkelTaylor}, the traceless quantity:
\begin{equation}
\label{PiBar}\bar \Pi ^{\mu \nu ,\alpha \beta }\left( k\right) \equiv \Pi
^{\mu \nu ,\alpha \beta }\left( k\right) -\frac 14\left( \eta ^{\mu \alpha
}\Pi ^\rho {}_{\rho ,}{}^{\nu \beta }+\eta ^{\mu \beta }\Pi ^\rho {}_{\rho
,}{}^{\nu \alpha }+\eta ^{\nu \alpha }\Pi ^\rho {}_{\rho ,}{}^{\mu \beta
}+\eta ^{\nu \beta }\Pi ^\rho {}_{\rho ,}{}^{\mu \alpha }\right) ,
\end{equation}
represents at high temperatures a quantity which does not depend on the
choice of basic graviton fields. In this domain, the masses can be
effectively neglected so the theory is invariant under conformal
transformations. The representation-independence of $\bar \Pi $ then follows
in consequence of the Ward and Weyl identities. As we have seen, the
contributions from internal massless particles, are invariant under local
coordinate and conformal transformations. Consequently, the physical
amplitude given by (\ref{PiBar}) can be identified in this case with the
graviton self-energy even at finite temperature.

However for contributions from thermal matter, which are characterized by
the presence of massive particles, $\bar \Pi ^{\mu \nu ,\,\alpha \beta }$ is
no longer independent of the graviton field parametrization at finite
temperatures. Our task is to generalize (\ref{PiBar}) in such a way that the
corresponding quantity should represent a physical graviton self-energy at
all temperatures. To achieve this we consider the effective action which
generates the one-particle irreducible thermal Green functions. In the
representation (\ref{gdown}) for the $h$ fields, we have that:
\begin{equation}
\label{SEff}S_{eff}=\Gamma _h^{\alpha \beta }h_{\alpha \beta }\left(
0\right) +\frac 12\int d^4k\Pi _h^{\mu \nu ,\,\alpha \beta }\left( k\right)
\,h_{\mu \nu }\left( k\right) \,h_{\alpha \beta }\left( -k\right) +\cdots .
\end{equation}
Starting from these fields, the most general re-parametrization of the
graviton fields can be written as:
\begin{equation}
\label{hPrime}h_{\prime }^{\mu \nu }=a\,h^{\mu \nu }+b\,h_{\;\;\lambda
}^\lambda \,\eta ^{\mu \nu }+c\,h_{\;\;\lambda }^\lambda \,h^{\mu \nu
}+d\,h^{\mu \lambda }\,h_{\;\;\lambda }^\nu +e\,\left( h_{\;\;\lambda
}^\lambda \right) ^2\,\eta ^{\mu \nu }+f\,h_{\alpha \beta }h^{\alpha \beta
}\,\eta ^{\mu \nu }+\cdots \;\;,
\end{equation}
where $a,$ $b,$ $c,$ $d,$ $e,$ and $f$ denote arbitrary constants. For
example, a basic graviton field often used in the literature
which is defined by \cite{FrenkelTaylorBrandt}
\begin{equation}
\label{gTilde}\sqrt{-g\left( x\right) }g^{\mu \nu }\equiv
\eta ^{\mu \nu }+\kappa
h_{\prime }^{\mu \nu },
\end{equation}
corresponds to a special case of (\ref{hPrime}), with:
\begin{equation}
\label{SpecialParameters}a=-1;\;\;b=\frac 12;\;\;c=-\frac
12;\;\;d=1;\;\;e=\frac 18;\;\;f=-\frac 14.
\end{equation}
In what follows we shall assume, without loss of generality that $a^2=1$.
This can always be achieved by a further rescaling of the $h_{\prime }$
fields in (\ref{hPrime}) [see ref. \cite{AlmeidaFrenkelTaylor}].
Furthermore, we shall consider for simplicity the class of
parametrizations characterized by the conditions
\begin{equation}
\label{constraints}a\,b-a\,c+2\,b\,d=0;\qquad b+2\,f=0,
\end{equation}
which are explicitly verified by all the graviton representations
discussed in the literature
\cite{GrossPerryYaffe,GriboskyDonoghue,Rebhan,FrenkelTaylorBrandt}.

Since the effective action is invariant under a general re-parametrization
of the graviton fields, it can be written in terms of $h_{\prime }$ as:
\begin{equation}
\label{SEffPrime}S_{eff}=\Gamma _{h_{\prime }}^{\alpha \beta }h_{\prime
\alpha \beta }\left( 0\right) +\frac 12\int d^4k\Pi _{h_{\prime }}^{\mu \nu
,\,\alpha \beta }\left( k\right) \,h_{\prime \mu \nu }\left( k\right)
\,h_{\prime \alpha \beta }\left( -k\right) +\cdots .
\end{equation}
Identifying with the help of (\ref{hPrime}), the corresponding terms in (%
\ref{SEff}) and (\ref{SEffPrime}), we obtain the following relations:
\begin{equation}
\label{GammahhPrime}\Gamma _h^{\mu \nu }=a\,\Gamma_{h_\prime}^{\mu \nu }+
b\, \eta^{\mu\nu} \Gamma_{h_\prime\;\rho}^{\rho},
\end{equation}
\begin{equation}
\label{PihhPrime}
\begin{array}{lll}
\Pi_h^{\mu \nu ,\,\alpha \beta } \left(k\right)
& =\Pi_{h_\prime }^{\mu \nu ,\,\alpha\beta } \left(k\right)
& +\, a\,b\left(\Pi_{h_\prime }^{\mu \nu ,\rho }\,_\rho
\ \eta^{\alpha \beta }+\Pi_{h_\prime }^{\alpha \beta ,\ \rho }\,_\rho
\ \eta ^{\mu\nu }\right)
+\, b^2 \,\Pi_{h_\prime\sigma\;\;\rho }^{\sigma\;\;,\,\rho }
\eta^{\mu\nu}\,\eta^{\alpha\beta}\\
&  & +\,
\displaystyle c\left( \Gamma _{h_\prime }^{\mu \nu }\eta ^{\alpha \beta}
+\Gamma _{h_\prime }^{\alpha \beta }\eta ^{\mu \nu }\right)  \\
 &  & +\,
\displaystyle\frac d2\left( \Gamma _{h_\prime }^{\mu \alpha }\eta ^{\nu \beta}
+\Gamma _{h_\prime }^{\nu \beta }\eta ^{\mu \alpha }+\Gamma _{h_\prime }^{\nu
\alpha }\eta ^{\mu \beta }+\Gamma _{h_\prime }^{\mu \beta }\eta ^{\nu \alpha}
\right) \\
& & +\,
2 e \,\Gamma_{h_\prime\;\rho}^{\rho} \eta^{\mu\nu} \eta^{\alpha\beta} +
f \, \Gamma_{h_\prime\;\rho}^{\rho} \left(\eta^{\mu\beta} \eta^{\nu\alpha}+
\eta^{\mu\alpha}\eta^{\nu\beta}\right)
\end{array}.
\end{equation}

When the theory is invariant under conformal transformations, the following
Weyl identity holds \cite{AlmeidaFrenkelTaylor}:
\begin{equation}
\label{trPiprime}
\Pi _{h_{\prime }\;\rho \,}^{\rho \;\;\;\;,\,\mu \nu }=-\kappa
\frac{a+4c+2d}{a\left( a+4b\right) }\;\Gamma _{h_{\prime }}^{\mu \nu }\equiv
-\tilde \Gamma _{h_{\prime }}^{\mu \nu },
\end{equation}
where $\tilde \Gamma _{h_{\prime }}^{\mu \nu }$ is a traceless function. In
general, this is no longer true in the presence of thermal matter at finite
temperature. In order to take this fact into account, we generalize (\ref
{PiBar}) by considering the following traceless quantity:
\begin{equation}
\label{PiStar}
\begin{array}{lll}
\tilde\Pi ^{\mu \nu ,\,\alpha \beta }\left( k\right)
&\equiv\bar \Pi^{\mu \nu ,\,\alpha \beta }\left( k\right)
+\frac 12\Delta_{\;\lambda}^\lambda &
\left(\eta^{\mu\alpha}\eta^{\nu\beta}+\eta^{\mu\beta}\eta^{\nu\alpha}\right)
- \,
\left(
\eta^{\mu\nu}\Delta^{\alpha\beta}+\eta^{\alpha\beta}\Delta^{\mu\nu}
\right) + \\
 & &
\left(
\eta^{\mu\alpha}\Delta^{\nu\beta}+\eta^{\mu\beta}\Delta^{\nu\alpha}
 +\,
\eta^{\nu\alpha}\Delta^{\mu\beta}+\eta^{\nu\beta}\Delta^{\mu\alpha}
\right)
\end{array},
\end{equation}
where the tensor $\Delta^{\mu\nu}$ is given by:
\begin{equation}
\label{Delta}\Delta ^{\mu \nu }=\frac 14\left(\Pi_{\;\rho }^{\rho \;,\mu\nu }+
\tilde \Gamma ^{\mu \nu }\right) -
{\frac 1{32}}\left(\Pi^{\rho\;\; ,\sigma}_{\;\;\rho\;\;\sigma}+
\tilde\Gamma^\rho_{\;\rho}\right)\,\eta^{\mu\nu} .
\end{equation}
We remark that when the Weyl identity (\ref{trPiprime}) is applicable,
$\Delta ^{\mu \nu }$ vanishes so that Eq. (\ref{PiStar}) reduces to
(\ref {PiBar}) as expected. For this reason, only the contributions
from thermal matter will appear in Eq. (\ref{Delta}).

It is now straightforward to verify, with the help of the relations
(\ref{constraints}), (\ref{GammahhPrime}) and (\ref{PihhPrime}) that:
\begin{equation}
\label{ParametrizationInvariance}
\tilde\Pi _h^{\mu \nu ,\,\alpha \beta }\left(k\right) =
\tilde\Pi _{h_{\prime }}^{\mu \nu ,\,\alpha \beta }\left(
k\right) .
\end{equation}
This equation shows that the graviton self-energy given by the
relations (\ref{PiStar}) and (\ref{Delta}) is invariant under
re-parametrizations of graviton fields at all temperatures. In order
to understand the mechanism which enforces the above property, using
the relations (\ref{PiBar}), (\ref{PiStar}) and (\ref{Delta}), we write
the expression for the graviton self-energy in the form:
\begin{equation}
\label{PiPiTad}\tilde\Pi^{\mu\nu ,\,\alpha\beta}\left(k\right)=
\Pi^{\mu\nu ,\,\alpha\beta}\left(k\right)+
\Pi_{tad}^{\mu\nu ,\, \alpha\beta}\left(k\right),
\end{equation}
where $\Pi_{tad}^{\mu\nu ,\,\alpha\beta}$ is given by:
\begin{equation}
\label{PiTad}
\begin{array}{lll}
\Pi_{tad}^{\mu\nu ,\,\alpha\beta}\left(k\right)\equiv\displaystyle
{\frac {1}{4}}&\left(\eta^{\mu\alpha}\tilde\Gamma^{\nu\beta}+
                         \eta^{\mu\beta} \tilde\Gamma^{\nu\alpha}+
                         \eta^{\nu\alpha}\tilde\Gamma^{\mu\beta}+
                         \eta^{\nu\beta}\tilde\Gamma^{\mu\alpha}\right)\\
&-\eta^{\mu\nu}\Delta^{\alpha\beta}-\eta^{\alpha\beta}\Delta^{\mu\nu}
\end{array}.
\end{equation}
As mentioned before, $\Delta$ vanishes in the case when the thermal
fields can be considered as being effectively massless. The
contributions associated with the graviton 1-point function in
(\ref{PiTad}) can be represented diagrammatically as shown in Fig.
\ref{fig5}. Both $\Pi$ and $\tilde\Gamma$ depend individually on the
choice of basic graviton fields in a way that ensures $\tilde\Pi$ to
be independent of these parametrizations. Hence, in order to obtain a
physical self-energy, one must consider in addition to 1PI graviton
2-point function, also the corresponding ``tadpole'' contributions.
Since the graviton self-energy
(\ref{PiStar}) is parametrization-independent, it may be conveniently
evaluated in the representation (\ref{gdown}) of the gravitational
fields, from the contributions of thermal matter and radiation fields
given respectively by Eqs. (\ref{PI1T}) and (\ref{PI2T}).

\section{The effective graviton propagator}
\label{sec6}

In order to investigate the thermal mass of gravitons, we will study
the properties of the poles in the effective graviton propagator. This
is obtained by iterative insertions of the physical self-energy in the
classical graviton propagator $k^{-2}P^{\mu\nu}_{\alpha\beta}$, where:

\begin{equation}
P^{\mu\nu}_{\alpha\beta}=\frac{\delta^\mu_\alpha\delta^\nu_\beta+
\delta^\mu_\beta\delta^\nu_\alpha-\eta^{\mu\nu}\eta_{\alpha\beta}}2
\equiv 1{\hskip-.35em}1^{\mu\nu}_{\alpha\beta}-
\frac{\eta^{\mu\nu}\eta_{\alpha\beta}}2
\end{equation}
which is insensitive to changes of
parametrizations\cite{GrossPerryYaffe,AlmeidaFrenkelTaylor}.

As we have seen, because of the inclusion of graviton 1-point
functions, the graviton self-energy (\ref{PiPiTad})
is also independent of the parametrizations of graviton fields. These
properties ensure that physical quantities such as masses are
independent of the choice of basic graviton fields.  Using the fact
that the physical self-energy is traceless, $P$ behaves
effectively like the identity when acting on $\tilde\Pi$.  Hence, the
effective graviton propagator can be written in the form:

\begin{equation}
\label{DTilde}
\tilde{D}_{\alpha \beta }^{\mu \nu }(k)=
\frac1{k^2}\left[P_{\alpha\beta}^{\mu\nu}+
\frac1{k^2}\tilde{\Pi}_{\alpha\beta}^{\mu\nu}(k)+
\left(\frac1{k^2}\right)^2\tilde{\Pi}_{\rho\sigma}^{\mu\nu}(k)
\tilde{\Pi}_{\alpha\beta}^{\rho\sigma}(k)+\dots\right]
\end{equation}
The right-hand side of this equation sums up to a geometric series,
giving the relation:

\begin{equation}
\label{eqdt}
\left(k^2P^{\mu\nu}_{\rho\sigma}-\tilde\Pi^{\mu\nu}_{\rho\sigma}\right)
\tilde D^{\rho\sigma}_{\alpha\beta}=1{\hskip-.35em}1^{\mu\nu}_{\alpha\beta}
\end{equation}

The effective propagator satisfies certain fundamental constrains.  In
view of the traceless property of $\tilde\Pi$, Eq. (\ref{DTilde})
requires that:

\begin{equation}
\tilde D^{\rho\;\; ,\mu\nu}_\rho=-\frac{\eta_{\mu\nu}}{k^2}
\end{equation}
Furthermore, the Ward identity (\ref{Ward}) expressing gauge invariance
requires a longitudinal contribution in $\tilde\Pi$ connected with the
background energy-momentum tensor [cf. Eq. (2.16) in the
second paper of Ref. \cite{AlmeidaFrenkelTaylor}].
Considering for definiteness the high-temperature limit,
it is then straightforward to verify that Eq. (\ref{DTilde}) implies:

\begin{equation}
k_\alpha k_\beta\tilde D^{\alpha\beta}_{\mu\nu}=
{\frac {k_\mu k_\nu}{k^2}}-\frac12\eta_{\mu\nu}+
{\frac{\kappa^2\rho}{12}}
(\eta^{\alpha\beta}-4\delta_0^\alpha\delta_0^\beta)
(k^21{\hskip-.35em}1-\tilde\Pi)^{-1}_{\alpha\beta,\,\mu\nu}
\end{equation}
where the energy density $\rho$ is given by:
\begin{equation}
\label{density}\rho=\frac{\omega\pi^2T^4}{30}.
\end{equation}
Here $\omega$ denotes the total number of degrees of freedom of the
thermal particles.

In what follows we shall be interested only in determining the
effective dynamical masses in the static case $k_0=0$, which are relevant in
the process of dynamical screening. To this end, we
project the corresponding contributions of $\tilde\Pi$ into the
following traceless normal modes:

\begin{eqnarray}
T_J^{\mu\nu,\,\alpha\beta}&=&\left[-\frac43 T_3(u)-\frac1{12}T_4(u)+
\frac13 T_5(u)\right]^{\mu\nu,\,\alpha\beta}\\
T_S^{\mu\nu,\,\alpha\beta}&=&
\left[-\frac12T_2(u)+2T_3(u)\right]^{\mu\nu,\,\alpha\beta},
\end{eqnarray}
where the tensors $T^{\mu\nu,\,\alpha\beta}_i(u)$
($i=1,\cdots,5$) are obtained from the corresponding ones in
Table \ref{table1} by replacing
$X^\alpha$ with $u^\alpha\equiv\delta^\alpha_0$.
The normal modes $T_J$ and
$T_S$ are idempotent (up to a minus sign)
and orthogonal to each other. While the mode $T_J$ is three-dimensionally
longitudinal, the mode $T_S$ is spatially-transverse in the sense that:

\begin{equation}
k_i k_j T_S^{i\, j,\,\,\mu\nu}=0;\qquad \left(i,\, j=1,2,3\right).
\end{equation}

In terms of these tensors, we can decompose the self-energy as follows:

\begin{equation}
\tilde\Pi^{\mu\nu,\,\alpha\beta}(k_0=0)=\kappa^2\rho
\left(T_J^{\mu\nu,\,\alpha\beta}-\frac13
T_S^{\mu\nu,\,\alpha\beta}\right)
\end{equation}
It is then easy to invert Eq. (\ref{eqdt}), yielding for the effective
graviton propagator the result:

\begin{equation}
\tilde D^{\mu\nu,\,\alpha\beta}\left(k_0=0\right)
=\frac1{\vec k^2}T_U^{\mu\nu,\,\alpha\beta}+
\frac1{\vec k^2+\frac{\kappa^2\rho}3}T_S^{\mu\nu,\,\alpha\beta}+
\frac1{\vec k^2-\kappa^2\rho}T_J^{\mu\nu,\,\alpha\beta}
\end{equation}
where the normal mode $T_U$ given by:

\begin{equation}
T_U^{\mu\nu,\,\alpha\beta}=\left[-\frac12T_1(u)+\frac12T_2(u)-\frac23T_3(u)+
\frac7{12}T_4(u)-\frac13T_5(u)\right]^{\mu\nu,\,\alpha\beta}
\end{equation}
is orthogonal to the modes $T_J$ and $T_S$.

We see that in the normal mode $T_U$, the gravitational plasma is
unscreened. This is somewhat similar to the spatially-transverse mode in the
QCD plasma. On the other hand, a non-vanishing
screening mass appears in the mode $T_S$:

\begin{equation}
m_S^2=\frac{\kappa^2\rho}3=\frac{32\pi G\rho}3
\end{equation}
analogously to the behavior shown by the spatially-longitudinal mode
in the QCD plasma.

The mode $T_J$ is characterized by  an imaginary mass:

\begin{equation}
m_J^2=-32\pi G\rho
\end{equation}
which is similar to the classical Jeans mass. This anti-screening
mode indicates a
gravitational instability for density fluctuations with wavelength
larger than $|m_J|^{-1}$, owing to the attractive nature of gravity.
One may generalize this calculation by including internal gravitons
in thermal equilibrium at high temperatures\cite{FrenkelTaylorBrandt}.
Their contributions will not affect the above
conclusion, since these change only the weight factor $w$ appearing in $\rho$
[cf. Eq.(\ref{density})] which counts the total number of degrees of freedom.

In conclusion we emphasize that this work, like that of
Ref.\cite{GrossPerryYaffe}, has been concerned with
gravitational perturbations
at finite temperatures around the Minkowski background, in an
asymptotically flat space. Using a different approach, based on the
study of small disturbances around the solutions of Einstein
equations, Rebhan\cite{Rebhan} has performed a rather complete
investigation of the gravitational instabilities. These metric
perturbations are relevant at high temperatures in the context of a
radiation dominated Robertson-Walker~universe.

\acknowledgements A.P.A and F.T.B. are grateful to FAPESP and CNP$_{\text{q}%
} $ (Brasil) for financial support. J.F. would like to thank Professor J.C.
Taylor for a helpful correspondence.

\appendix

\newpage
\section{}
\label{appA}

In this appendix we present the Feynman rules for the couplings and
propagators involving scalar, gauge and graviton fields. These rules
can be obtained from the respective
Lagrangians given in Eq. (\ref{MatterLagrangian})
and Eqs. (\ref{RadiationLagrangian}), (\ref{GaugeFix}) and (\ref{Ghost}).
In a perturbative calculation we first have to expand all the metric dependent
quantities up to some given order in the graviton field $h$.
These expansions and the subsequent
reading of the momentum space Feynman rules is a straightforward procedure
(but a very tedious task for humans) which was accomplished using a
algebraic computer algorithm written in Mathematica.
Here we will only
present the results for the vertices involving up to two gravitons, which are
relevant for the calculation of the graviton polarization tensor. We will also
restrict only to the Abelian couplings of the gauge fields
[cf. Eq. (\ref{RadiationLagrangian})].

In all the expressions which follows we will always denote the graviton
momenta and indices by  $[k_1,\,(\mu,\,\nu)]$ and $[k_2,\,\alpha,\,\beta)]$.
The momenta of scalars and ghosts are denoted by $p_1$ and $p_2$. The gluon
momenta and indices are denoted by $[p_1,\,\rho]$ and $[p_2,\,\sigma]$.
Using this notation, we obtain from the Lagrangian (\ref{MatterLagrangian})
the {\it scalar-scalar-graviton} interaction vertex
\begin{equation}
\label{v1scal}
\displaystyle{\frac {2}{\kappa}} V^{^1{scalar}}_{\mu\nu}(k_1;p_1,p_2)=
\,p_{_1\mu }\,p_{_2\nu } + \,p_{_1\nu }\,p_{_2\mu }  -
  p_1 \cdot p_2\,\eta_{\mu  \nu } - {m^2}\,\eta_{\mu  \nu } +
2\, \xi \,\left( \,k_{_1\mu }\,k_{_1\nu } - \,k_1^{2}\,\eta_{\mu  \nu } \right)
\end{equation}
and the {\it scalar-scalar-graviton-graviton} interaction vertex
\begin{equation}
\begin{array}{lll}
\label{v2scal}
\displaystyle{\frac {16}{\kappa^2}}
&V^{^2{scalar}}_{\mu\nu,\,\alpha\beta}(k_1,k_2;p_1,p_2)=&
-8\,p_{_1\nu }\,p_{_2\beta }\,\eta_{\alpha  \mu } +
  2\,p_1 \cdot p_2\,\eta_{\alpha  \mu }\,\eta_{\beta  \nu } +
  4\,p_{_1\alpha }\,p_{_2\beta }\,\eta_{\mu  \nu } - \\
&&  p_1 \cdot p_2\,\eta_{\alpha  \beta }\,\eta_{\mu  \nu } +
   {m^2}\,\left( 2\,\eta_{\alpha  \mu }\,\eta_{\beta  \nu } -
     \eta_{\alpha  \beta }\,\eta_{\mu  \nu } \right)  + \\
&&   \xi \,\left( 4\,k_{_1\mu }\,k_{_1\nu }\,\eta_{\alpha  \beta } +
     4\,k_{_1\mu }\,k_{_2\nu }\,\eta_{\alpha  \beta } -
     4\,k_{_1\beta }\,k_{_1\nu }\,\eta_{\alpha  \mu } - \right. \\
&&     6\,k_{_1\beta }\,k_{_2\nu }\,\eta_{\alpha  \mu } -
     4\,k_{_1\beta }\,k_{_1\mu }\,\eta_{\alpha  \nu } +
     2\,k_{_1\beta }\,k_{_2\mu }\,\eta_{\alpha  \nu } - \\
&&     4\,k_{_1\alpha }\,k_{_1\nu }\,\eta_{\beta  \mu } -
     2\,k_{_1\alpha }\,k_{_2\nu }\,\eta_{\beta  \mu } +
     2\,k_1 \cdot k_2\,\eta_{\alpha  \nu }\,\eta_{\beta  \mu } - \\
&&     4\,k_{_1\alpha }\,k_{_1\mu }\,\eta_{\beta  \nu } -
     8\,k_{_1\mu }\,k_{_2\alpha }\,\eta_{\beta  \nu } +
     2\,k_{_1\alpha }\,k_{_2\mu }\,\eta_{\beta  \nu } + \\
&&     8\,k_1^{2}\,\eta_{\alpha  \mu }\,\eta_{\beta  \nu } +
     4\,k_1 \cdot k_2\,\eta_{\alpha  \mu }\,\eta_{\beta  \nu } +
     8\,k_{_1\alpha }\,k_{_1\beta }\,\eta_{\mu  \nu } + \\
&& \left.    4\,k_{_1\beta }\,k_{_2\alpha }\,\eta_{\mu  \nu } -
       4\,k_1^{2}\,\eta_{\alpha  \beta }\,\eta_{\mu  \nu } -
     2\,k_1 \cdot k_2\,\eta_{\alpha  \beta }\,\eta_{\mu  \nu } \right),
\end{array}
\end{equation}
where in the expression above one has to perform a
symmetrization over the graviton indices and permutation of the
scalar particles.

{}From the ghost Lagrangian we obtain the
{\it ghost-ghost-graviton} interaction vertex
\begin{equation}
\label{ghost}
\displaystyle{\frac {2}{\kappa}} V^{^1ghost}_{\mu\nu}(k_1;\,p_1,p_2)=
-p_{_1\nu }\,p_{_2\mu }  -
  p_{_1\mu }\,p_{_2\nu } +
  p_1 \cdot p_2\,\eta_{\mu  \nu },
\end{equation}
and the {\it ghost-ghost-graviton-graviton} interaction vertex
\begin{equation}
\label{ghost2}
\displaystyle{\frac {8}{\kappa}}
V^{^2ghost}_{\mu\nu,\,\alpha\beta}(k_1,k_2;\,p_1,p_2)=
-4\,p_{_1\mu }\,p_{_2\nu }\,\eta_{\alpha  \beta } +
  8\,p_{_1\beta }\,p_{_2\nu }\,\eta_{\alpha  \mu } -
  2\,p_1 \cdot p_2\,\eta_{\alpha  \mu }\,\eta_{\beta  \nu } +
  p_1 \cdot p_2\,\eta_{\alpha  \beta }\,\eta_{\mu  \nu }.
\end{equation}
Expression (\ref{ghost2}) has to be symmetrized over the graviton indices.
Notice that, as can be
easily seen from Eqs. (\ref{MatterLagrangian}) and (\ref{Ghost}),
the  interaction vertices of ghosts or scalars with the graviton field,
differ only by a minus sign when $\xi=m=0$. The corresponding
propagators are given by
\begin{equation}
D(k)=\left\{\begin{array}{lll}&{\;\;
\displaystyle\frac{1}{k^2}}&{\rm for\;the\;ghost,}\\
& \displaystyle{\frac{1}{m^2-k^2}}\qquad &{\rm for\;the\;scalar}
            \end{array}
\right. \;\; .
\end{equation}

{}From Eqs. (\ref{RadiationLagrangian}) and (\ref{GaugeFix}) we can
obtain the gauge fields Feynman rules in the general covariant gauge
characterized by the gauge fixing parameter $\alpha$.
In this class of gauges the gauge field propagator is given by
\begin{equation}
D_{\mu\nu}(k)={\frac{1}{k^2}}\left[\eta_{\mu\nu}-
\left(1-\alpha\right){\frac{k_\mu\,k_\nu}{k^2}}\right].
\end{equation}
The interaction vertices will also depend on the parameter $\alpha$.
The {\it gauge-gauge-graviton} coupling is
\begin{equation}
\label{v1gauge}
\begin{array}{lll}
&\displaystyle{\frac {4}{\kappa}}
V^{^1{gauge}}_{\mu\nu;\rho,\,\sigma}(k_1;p_1,p_2)=&
-p_{_1\sigma }\,p_{_2\rho }\,\eta_{\mu  \nu }  +
  p_{_1\sigma }\,p_{_2\nu }\,\eta_{\mu  \rho } +
  p_{_1\nu }\,p_{_2\rho }\,\eta_{\mu  \sigma } + \\
&&  p_{_1\sigma }\,p_{_2\mu }\,\eta_{\nu  \rho } +
  p_{_1\mu }\,p_{_2\rho }\,\eta_{\nu  \sigma } -
  2\,p_1 \cdot p_2\,\eta_{\mu  \rho }\,\eta_{\nu  \sigma } - \\
&&  2\,p_{_1\mu }\,p_{_2\nu }\,\eta_{\rho  \sigma } +
  p_1 \cdot p_2\,\eta_{\mu  \nu }\,\eta_{\rho  \sigma }+ \\
&&  \displaystyle{\frac {1}{\alpha}}\left(
  -2\,p_{_1\rho }\,p_{_1\sigma }\,\eta_{\mu  \nu } -
      p_{_1\rho }\,p_{_2\sigma }\,\eta_{\mu  \nu } -
      2\,p_{_1\nu }\,p_{_2\sigma }\,\eta_{\mu  \rho } + \right. \\
&& \left.      4\,p_{_1\nu }\,p_{_1\rho }\,\eta_{\mu  \sigma } +
      2\,p_{_1\rho }\,p_{_2\nu }\,\eta_{\mu  \sigma }\right) ,
\end{array}
\end{equation}
and the {\it gauge-gauge-graviton-graviton} vertex is
\begin{equation}
\begin{array}{lll}
&\displaystyle{\frac {16}{\kappa^2}}
V^{^2{gauge}}_{\mu\nu,\,\alpha\beta;\,\rho,\,\sigma}(k_1,k_2;p_1,p_2)=&
2\,p_{_1\sigma}\,p_{_2\rho}\,\eta_{\alpha  \mu }\,
   \eta_{\beta  \nu } - 4\,p_{_1\sigma}\,p_{_2\nu }\,
   \eta_{\alpha  \mu }\,\eta_{\beta  \rho} -
  4\,p_{_1\nu }\,p_{_2\rho}\,\eta_{\alpha  \mu }\,
   \eta_{\beta  \sigma} + \\
&& 4\,p_{_1\mu }\,p_{_2\nu }\,
   \eta_{\alpha  \rho}\,\eta_{\beta  \sigma} -
  p_{_1\sigma}\,p_{_2\rho}\,\eta_{\alpha  \beta }\,
   \eta_{\mu  \nu } + 2\,p_{_1\sigma}\,p_{_2\beta }\,
   \eta_{\alpha  \rho}\,\eta_{\mu  \nu } + \\
&&  2\,p_{_1\beta }\,p_{_2\rho}\,\eta_{\alpha  \sigma}\,
   \eta_{\mu  \nu } + 2\,p_{_1\sigma}\,p_{_2\alpha }\,
   \eta_{\beta  \rho}\,\eta_{\mu  \nu } +
  2\,p_{_1\alpha }\,p_{_2\rho}\,\eta_{\beta  \sigma}\,
   \eta_{\mu  \nu } - \\
&& 4\,p_1 \cdot p_2\,\eta_{\alpha  \rho}\,
   \eta_{\beta  \sigma}\,\eta_{\mu  \nu } -
  4\,p_{_1\alpha }\,p_{_2\nu }\,\eta_{\beta  \sigma}\,
   \eta_{\mu  \rho}  -
  4\,p_{_1\sigma}\,p_{_2\beta }\,\eta_{\alpha  \mu }\,
   \eta_{\nu  \rho} - \\
&& 4\,p_{_1\beta }\,p_{_2\mu }\,
   \eta_{\alpha  \sigma}\,\eta_{\nu  \rho} +
  8\,p_1 \cdot p_2\,\eta_{\alpha  \mu }\,\eta_{\beta  \sigma}\,
   \eta_{\nu  \rho} - 4\,p_{_1\beta }\,p_{_2\rho}\,
   \eta_{\alpha  \mu }\,\eta_{\nu  \sigma} + \\
&&  4\,p_{_1\alpha }\,p_{_2\beta }\,\eta_{\mu  \rho}\,
   \eta_{\nu  \sigma} + 8\,p_{_1\nu }\,p_{_2\beta }\,
   \eta_{\alpha  \mu }\,\eta_{\rho \sigma} -
  2\,p_1 \cdot p_2\,\eta_{\alpha  \mu }\,\eta_{\beta  \nu }\,
   \eta_{\rho \sigma} - \\
&& 4\,p_{_1\alpha }\,p_{_2\beta }\,
   \eta_{\mu  \nu }\,\eta_{\rho \sigma} +
  p_1 \cdot p_2\,\eta_{\alpha  \beta }\,\eta_{\mu  \nu }\,
   \eta_{\rho \sigma} + \\
&& \displaystyle{\frac{1}{\alpha}}\left(
      16\,k_{_1\nu }\,p_{_1\rho}\,
       \eta_{\alpha  \sigma}\,\eta_{\beta  \mu } -
      8\,k_{_1\sigma}\,p_{_1\rho}\,\eta_{\alpha  \mu }\,
       \eta_{\beta  \nu } - 2\,p_{_1\rho}\,p_{_2\sigma}\,
       \eta_{\alpha  \mu }\,\eta_{\beta  \nu } +  \right. \\
&&      2\,k_{_1\rho}\,k_{_2\sigma}\,\eta_{\alpha  \beta }\,
       \eta_{\mu  \nu } + 4\,k_{_1\sigma}\,p_{_1\rho}\,
       \eta_{\alpha  \beta }\,\eta_{\mu  \nu } +
      p_{_1\rho}\,p_{_2\sigma}\,\eta_{\alpha  \beta }\,
       \eta_{\mu  \nu } - \\
&& 8\,k_{_1\sigma}\,p_{_1\beta }\,
       \eta_{\alpha  \rho}\,\eta_{\mu  \nu } -
      4\,p_{_1\beta }\,p_{_2\sigma}\,\eta_{\alpha  \rho}\,
       \eta_{\mu  \nu } - 4\,k_{_1\rho}\,k_{_2\beta }\,
       \eta_{\alpha  \sigma}\,\eta_{\mu  \nu } - \\
&&      8\,k_{_1\beta }\,p_{_1\rho}\,\eta_{\alpha  \sigma}\,
       \eta_{\mu  \nu } - 4\,p_{_1\rho}\,p_{_2\beta }\,
       \eta_{\alpha  \sigma}\,\eta_{\mu  \nu } -
      4\,k_{_1\nu }\,k_{_2\sigma}\,\eta_{\alpha  \beta }\,
       \eta_{\mu  \rho} + \\
&& 8\,p_{_1\beta }\,p_{_2\sigma}\,
       \eta_{\alpha  \nu }\,\eta_{\mu  \rho} +
      8\,k_{_1\nu }\,k_{_2\beta }\,\eta_{\alpha  \sigma}\,
       \eta_{\mu  \rho} + 8\,p_{_1\nu }\,p_{_2\beta }\,
       \eta_{\alpha  \sigma}\,\eta_{\mu  \rho} - \\
&&      8\,k_{_1\nu }\,p_{_1\rho}\,\eta_{\alpha  \beta }\,
       \eta_{\mu  \sigma} + 8\,k_{_1\beta }\,p_{_1\rho}\,
       \eta_{\alpha  \nu }\,\eta_{\mu  \sigma} +
      8\,p_{_1\rho}\,p_{_2\beta }\,\eta_{\alpha  \nu }\,
       \eta_{\mu  \sigma} + \\
&& \left. 16\,k_{_1\nu }\,p_{_1\beta }\,
       \eta_{\alpha  \rho}\,\eta_{\mu  \sigma} +
      8\,k_{_1\alpha }\,p_{_1\rho}\,\eta_{\beta  \nu }\,
       \eta_{\mu  \sigma}\right) .
\end{array}
\end{equation}
Similarly as in the scalar vertices, one has to symmetrize the
gauge field vertices
over the graviton indices and include the permutations of the gluons.
In all interaction vertices there is momentum conservation, with
all momenta inwards.

{}From the expressions above we can perform the explicit computation
of the scattering amplitudes shown in the figures \ref{fig2} and
\ref{fig4}. This was done using these Feynman rules as an input to a
algebraic computer program.

\section{}
\label{appB}
In this appendix we present the leading and next-to-leading structure
functions for the matter contribution to the polarization tensor.
The leading structure functions
presented here are the same for all thermal particles.
The explicit result for the functions
$l_i(r,K)$, $n_i(r,K)$ and $s_i(r,K)$ appearing in Eq. (\ref{CoefPI1})
is the following:
\begin{equation}\begin{array}{ll}
&l\,_1={\frac 16}-{\frac{{K^2}}{{24}}}+{\frac{{{K^4}\,L}}8}  \\
&n\,_1={\frac{{-5\,{K^4}}}{{192}}}-{\frac{{5\,{K^2}\,\xi }}{{16}}}+{\frac{{5\,%
{K^6}\,L}}{{64}}}  \\
&s\,_1=-{\frac 58}+{\frac{{5\,{K^2}\,L}}{{16}}}
\end{array}\;\; ,\end{equation}
\begin{equation}\begin{array}{ll}
&l\,_2=-{\frac 13}+{\frac{{K^2}}{{12}}}-{\frac{{5\,{K^4}}}{{24}}}+{\frac{{5\,{%
K^6}\,L}}8}  \\
&n\,_2={\frac{{-25\,{K^6}}}{{192}}}+{\frac{{5\,{K^6}\,L}}{{32}}}+{\frac{{25\,{%
K^8}\,L}}{{64}}}  \\
&s\,_2={\frac 58}-{\frac{{5\,{K^2}}}{{16}}}+{\frac{{15\,{K^4}\,L}}{{16}}}
\end{array}\;\; ,\end{equation}
\begin{equation}\begin{array}{ll}
&l\,_3={\frac{{-{K^2}}}3}+{\frac{{7\,{K^4}}}{{12}}}-{\frac{{35\,{K^6}}}{{24}}}%
+{\frac{{35\,{K^8}\,L}}8}  \\
&n\,_3={\frac{{-5\,{K^6}}}{{32}}}-{\frac{{175\,{K^8}}}{{192}}}+{\frac{{25\,{%
K^8}\,L}}{{16}}}+{\frac{{175\,{K^{10}}\,L}}{{64}}}  \\
&s\,_3={\frac{{5\,{K^2}}}8}-{\frac{{25\,{K^4}}}{{16}}}+{\frac{{75\,{K^6}\,L}}{%
{16}}}
\end{array}\;\; ,\end{equation}
\begin{equation}\begin{array}{ll}
&l\,_4={\frac{{-{K^2}}}{{24}}}+{\frac{{{K^4}\,L}}8}  \\
&n\,_4={\frac{{-5\,{K^2}}}{{32}}}-{\frac{{5\,{K^4}}}{{192}}}-{\frac{{5\,{K^2}%
\,\xi }}8}+{\frac{{5\,{K^4}\,L}}{{16}}}+{\frac{{5\,{K^6}\,L}}{{64}}}+{\frac{{%
5\,{K^4}\,\xi \,L}}4}  \\
&s\,_4={\frac{{5\,{K^2}\,L}}{{16}}}
\end{array}\;\; ,\end{equation}
\begin{equation}\begin{array}{ll}
&l\,_5={\frac{{K^2}}{{12}}}-{\frac{{5\,{K^4}}}{{24}}}+{\frac{{5\,{K^6}\,L}}8}
\\
&n\,_5={\frac{{-5\,{K^4}}}{{32}}}-{\frac{{25\,{K^6}}}{{192}}}-{\frac{{5\,{K^4}%
\,\xi }}8}+{\frac{{5\,{K^6}\,L}}8}+{\frac{{25\,{K^8}\,L}}{{64}}}+{\frac{{15\,%
{K^6}\,\xi \,L}}8}  \\
&s\,_5={\frac{{-5\,{K^2}}}{{16}}}+{\frac{{15\,{K^4}\,L}}{{16}}}
\end{array}\;\; ,\end{equation}
\begin{equation}\begin{array}{ll}
&l\,_6={\left( {\frac{{-1}}{{12}}}+{\frac{{5\,{K^2}}}{{24}}}-{\frac{{5\,{K^4}%
\,L}}8}\right) r}  \\
&n\,_6={\left( {\frac{{25\,{K^4}}}{{192}}}-{\frac{{5\,{K^4}\,L}}{{32}}}-{
\frac{{25\,{K^6}\,L}}{{64}}}\right) r}  \\
&s\,_6={\left( {\frac{{5}}{{16}}}-{\frac{{15\,{K^2}\,L}}{{16}}}\right) r}
\end{array}\;\; ,\end{equation}
\begin{equation}\begin{array}{ll}
&l\,_7={\left( {\frac 13}-{\frac{{7\,{K^2}\,}}{{12}}}+{\frac{{35\,{K^4}}}{{24}%
}}-{\frac{{35\,{K^6}\,L}}8}\right) r}  \\
&n\,_7={\left( {\frac{{5\,{K^4}}}{{32}}}+{\frac{{175\,{K^6}}}{{192}}}-{\frac{{%
25\,{K^6}\,L}}{{16}}}-{\frac{{175\,{K^8}\,L}}{{64}}}\right) r}  \\
&s\,_7={\left( {\frac{{-5}}8}+{\frac{{25\,{K^2}}}{{16}}}-{\frac{{75\,{K^4}\,L}
}{{16}}}\right) r}
\end{array}\;\; ,\end{equation}
\begin{equation}\begin{array}{ll}
&l\,_8=-{\frac 1{{12}}}-{\frac{{5\,{K^2}}}{{24}}}+{\frac{{{K^2}\,L}}2}+{\frac{%
{5\,{K^4}\,L}}8}  \\
&n\,_8={\frac{{-5\,{K^2}}}{{48}}}-{\frac{{25\,{K^4}}}{{192}}}+{\frac{{5\,\xi }
}{{16}}}+{\frac{{5\,{K^2}\,L}}{{32}}}+{\frac{{15\,{K^4}\,L}}{{32}}}+{\frac{{%
25\,{K^6}\,L}}{{64}}}  \\
&s\,_8=-{\frac 5{{16}}}+{\frac{{5\,L}}8}+{\frac{{15\,{K^2}\,L}}{{16}}}
\end{array}\;\; ,\end{equation}
\begin{equation}\begin{array}{ll}
&l\,_9={\frac 16}-{\frac{{2\,{K^2}}}3}-{\frac{{35\,{K^4}}}{{24}}}+{\frac{{15\,%
{K^4}\,L}}4}+{\frac{{35\,{K^6}\,L}}8}  \\
&n\,_9={\frac{{-15\,{K^4}}}{{16}}}-{\frac{{175\,{K^6}}}{{192}}}+{\frac{{5\,{%
K^2}\,\xi }}8}+{\frac{{15\,{K^4}\,L}}{{16}}}+{\frac{{125\,{K^6}\,L}}{{32}}}+{
\frac{{175\,{K^8}\,L}}{{64}}}-{\frac{{15\,{K^4}\,\xi \,L}}8}  \\
&s\,_9=-{\frac 58}-{\frac{{25\,{K^2}}}{{16}}}+{\frac{{15\,{K^2}\,L}}4}+{\frac{%
{75\,{K^4}\,L}}{{16}}}
\end{array}\;\; ,\end{equation}
\begin{equation}\begin{array}{ll}
&l\,_{10}={\frac 16}-{\frac{{2\,{K^2}}}3}-{\frac{{35\,{K^4}}}{{24}}}+{\frac{{%
15\,{K^4}\,L}}4}+{\frac{{35\,{K^6}\,L}}8}  \\
&n\,_{10}={\frac{{-5\,{K^2}}}{{32}}}-{\frac{{15\,{K^4}}}{{16}}}-{\frac{{175\,{%
K^6}}}{{192}}}+{\frac{{45\,{K^4}\,L}}{{32}}}+{\frac{{125\,{K^6}\,L}}{{32}}}+{
\frac{{175\,{K^8}\,L}}{{64}}}  \\
&s\,_{10}=-{\frac 58}-{\frac{{25\,{K^2}}}{{16}}}+{\frac{{15\,{K^2}\,L}}4}+{
\frac{{75\,{K^4}\,L}}{{16}}}
\end{array}\;\; ,\end{equation}
\begin{equation}\begin{array}{ll}
&l\,_{11}={\left( {\frac 14}+{\frac{{35\,{K^2}}}{{24}}}-{\frac{{5\,{K^2}\,L}}2%
}-{\frac{{35\,{K^4}\,L}}8}\right) r}  \\
&n\,_{11}={\left( {\frac{{65\,{K^2}}}{{96}}}+{\frac{{175\,{K^4}}}{{192}}}-{
\frac{{5\,\xi }}8}-{\frac{{5\,{K^2}\,L}}8}-{\frac{{25\,{K^4}\,L}}8}-{\frac{{%
175\,{K^6}\,\,L}}{{64}}}+{\frac{{15\,{K^2}\,\xi \,L}}8}\right) r}  \\
&s\,_{11}={\left( {\frac{{25}}{{16}}}-{\frac{{15\,L}}8}-{\frac{{75\,{K^2}\,L}
}{{16}}}\right) r}
\end{array}\;\; ,\end{equation}
\begin{equation}\begin{array}{ll}
&l\,_{12}=-{\frac{13}{{12}}}-{\frac{{35\,{K^2}}}{{24}}}+L+5\,{K^2}\,L+{\frac{{%
35\,{K^4}\,L}}8}  \\
&n\,_{12}=-{\frac 5{{24}}}-{\frac{{115\,{K^2}}}{{96}}}-{\frac{{175\,{K^4}}}{{%
192}}}+{\frac{{5\,\xi }}4}+{\frac{{15\,{K^2}\,L}}8}+{\frac{{75\,{K^4}\,L}}{{%
16}}}+{\frac{{175\,{K^6}\,L}}{{64}}}-{\frac{{5\,\xi \,L}}2}-{\frac{{15\,{K^2}%
\,\xi \,L}}4}  \\
&s\,_{12}=-{\frac{25}{{16}}}-{\frac 5{{8\,{K^2}}}}+{\frac{{15\,L}}4}+{\frac{{%
75\,{K^2}\,L}}{{16}}}
\end{array}\;\; ,\end{equation}
\begin{equation}\begin{array}{ll}
&l\,_{13}=-{\frac 1{{12}}}-{\frac{{5\,{K^2}}}{{24}}}+{\frac{{{K^2}\,L}}2}+{
\frac{{5\,{K^4}\,L}}8}  \\
&n\,_{13}={\frac{{-25\,{K^2}}}{{96}}}-{\frac{{25\,{K^4}}}{{192}}}-{\frac{{5\,{%
K^2}\,\xi }}8}+{\frac{{5\,{K^2}\,L}}{{16}}}+{\frac{{15\,{K^4}\,L}}{{16}}}+{
\frac{{25\,{K^6}\,L}}{{64}}}+{\frac{{5\,{K^2}\,\xi \,L}}8}+{\frac{{15\,{K^4}%
\,\xi \,L}}8}  \\
&s\,_{13}=-{\frac 5{{16}}}+{\frac{{5\,L}}8}+{\frac{{15\,{K^2}\,L}}{{16}}}
\end{array}\;\; ,\end{equation}
\begin{equation}\begin{array}{ll}
&l\,_{14}={\left( {\frac{{-1}}{{12}}}+{\frac{{5\,{K^2}}}{{24}}}-{\frac{{5\,{%
K^4}\,L}}8}\right) r}  \\
&n\,_{14}={\left( {\frac{{5\,{K^2}}}{{32}}}+{\frac{{25\,{K^4}}}{{192}}}+{
\frac{{5\,{K^2}\,\xi }}8}-{\frac{{5\,{K^4}\,L}}8}-{\frac{{25\,{K^6}\,L}}{{64}%
}}-{\frac{{15\,{K^4}\,\xi \,L}}8}\right) r}  \\
&s\,_{14}={\left( {\frac{{5}}{{16}}}-{\frac{{15\,{K^2}\,L}}{{16}}}\right) r}
\end{array}\;\; .\end{equation}
The dimensionless quantity $L$ is a function of r given by:
\begin{equation}
\label{Lfun}
L\left( r\right) =\frac r2\ln \frac{r+1}{r-1}-1.
\end{equation}

\section{}
\label{appC}

Here we calculate the integrals

\begin{equation}
\label{c1}
J_n(t)=\left(\frac{i\pi T}{|\vec k|}\right)^{n-1}\int^tdy\,y^n
\left[\ln(y)-\frac1{2y}-\psi(y)\right]
\end{equation}
in terms of the generalized $\zeta$ function, defined as

\begin{equation}
\label{c2}
\zeta(z,y)=\sum_{l=0}^\infty\frac1{(l+y)^z}.
\end{equation}

To this end we express the $\psi$ function as

\begin{equation}
\label{c5}
\psi(y)=\lim_{\varepsilon\rightarrow0}\left[\frac1\varepsilon-
\zeta(1+\varepsilon,y)
\right]
\end{equation}
and use the formula

\begin{equation}
\label{c3}
\int^tdy\zeta(z,y)=\frac1{1-z}\zeta(1-z,t),
\end{equation}
that can be easily verified from eq. (\ref{c2}), and can be
generalized to

\begin{equation}
\label{c4}
\int^tdy y^n\zeta(z,y)=\sum^n_{l=0}\frac{(n-l)!}{n!}
\frac{\Gamma(1-z)}{\Gamma(l+2-z)}t^n\zeta(z-l-1,t),
\end{equation}
which is eq. (\ref{c3}) integrated by parts $n$ times.

Substituting eq. (\ref{c5}) in eq. (\ref{c1}) and using eq. (\ref{c4})
we can verify, using the properties of $\zeta$ function, that the
divergent term as $\varepsilon\rightarrow0$ cancels out, as
expected. For $n\neq0$ the remaining terms give:
\begin{eqnarray}
\label{eqj}
J_n(t)&=&\left(\frac{i\pi T}{|\vec k|}\right)^{n-1}
\left\{\frac{t^{n+1}}{n+1}\ln t-\frac{t^{n+1}}{(n+1)^2}-\frac{t^{n}}{2n}-
\sum_{j=0}^n\left( -1 \right)^jt^{n-j}{n\choose j}\zeta'(-j,t)+
\right.
\nonumber\\
&&\left.\sum_{j=1}^n\frac{\left(-1\right)^j}{j+1}{n\choose j}
\left(\sum_{k=1}^j\frac1k\right)t^{n-j}{\rm B}_{j+1}(t)\right\},
\end{eqnarray}
where ${\rm B}_n$ are the Bernoulli polynomials\cite{Gradshteyn}. For
$n=0$, $J_0(t)$ is given by Eq. (\ref{ExpampleOfJ}).

Now we discuss the behavior of the generalized zeta function for
asymptotic values of the parameter $t(k_0)=i\frac{k_0+|{\bf k}|}{4\pi
T}$, which correspond to high temperature expansion. To this end, we
start from the representation\cite{Gradshteyn}:

\begin{equation}
\label{defz}
\zeta(z,t)=\frac1{t^z}+
\frac1{\Gamma(z)}\int^\infty_0\frac{x^{z-1}e^{-tx}}{e^x-1}dx.
\end{equation}

Expanding (\ref{defz}) in power series of $t$, making use of the
integral representation of Riemann's zeta function and Euler's gamma
function we find

\begin{equation}
\label{expaz}
\zeta(z,t)=\frac1{t^z}+
\sum_{l=0}^\infty\frac{\Gamma(z+l)}{\Gamma(z)}\frac{(-t)^l}{l!}
\zeta(z+l).
\end{equation}

Taking the derivative of (\ref{expaz}) with respect to $z$ we obtain
in terms of the psi function $\psi(z)$ that

\begin{equation}
\zeta'(z,t)=\sum_{l=0}^\infty\frac{(-t)^l}{l!}\frac{\Gamma(z+l)}{\Gamma(z)}
\left\{\left[\psi(z+l)-\psi(z)\right]\zeta(z+l)+\zeta'(z+l)\right\}-
t^z {\rm ln}(z).
\end{equation}

We are actually interested in the values of $\zeta'(z,t)$ for
$z\rightarrow-n$ where $n$ is a natural number. After a
straightforward calculation we obtain

\begin{eqnarray}
\label{expzl}
\zeta'(-n,t)&=&\sum\limits_{l=0}^n{n\choose l}\left[\zeta'(l-n)-
\zeta(l-n)\sum\limits_{k=n-l+1}^n\frac1k\right]t^l-t^n{\rm ln}(z)\nonumber\\
&&
-\frac{t^{n+1}}{n+1}\left(\gamma-\sum_{k=1}^n\frac1k\right)
+\sum\limits_{l=n+2}^\infty(-1)^{n+1}t^l\frac{n!(l-n-1)!}{l!}\zeta(l-n).
\end{eqnarray}

With help of this
formula we can compute the functions $J_n$ in Eq.
(\ref{eqj}) and express $P_n$ from Eqs.  (\ref{p1}), (\ref{p2}) and
(\ref{p3}) as a series of decreasing powers of $T$. Then, it is
straightforward to arrive at Eqs.  (\ref{p1t2}), (\ref{p2t2}) and
(\ref{p3t2}).

\vfill\eject

\begin{table}
  \begin{tabular}{lcc} \hline
$T^{\mu\nu ,\, \alpha\beta}_1(X,Y)=\eta ^{\alpha \nu }\,\eta ^{\beta \mu }+
\eta ^{\alpha \mu }\,\eta ^{\beta \nu }$  \\
$T^{\mu\nu ,\, \alpha\beta}_2(X,Y)=X^\mu \,\left( X^\beta
\,\eta ^{\alpha \nu }+X^\alpha \,\eta ^{\beta \nu }\right) +
X^\nu \,\left( X^\beta \,\eta ^{\alpha \mu }+X^\alpha \,\eta ^{\beta \mu }
\right) $ \\
$T^{\mu\nu ,\, \alpha\beta}_3(X,Y)=X^\alpha \,X^\beta \,X^\mu \,X^\nu $ \\
$T^{\mu\nu ,\, \alpha\beta}_4(X,Y)=\eta ^{\alpha \beta }\,\eta ^{\mu \nu }$  \\
$T^{\mu\nu ,\, \alpha\beta}_5(X,Y)=X^\mu \,X^\nu \,\eta ^{\alpha \beta }+
X^\alpha \,X^\beta \,\eta ^{\mu \nu }$ \\
$T^{\mu\nu ,\, \alpha\beta}_6(X,Y)=X^\beta \,\left( Y^\nu \,
\eta ^{\alpha \mu }+Y^\mu \,\eta ^{\alpha \nu }\right) +
Y^\beta \,\left( X^\nu \,\eta ^{\alpha \mu }+X^\mu \,\eta ^{\alpha \nu }
\right) $ \\
$\;\;\;\;\;\;\;\;\,\,\,\,\,\;\;\;\;\;\;\;\;\;\;\;
+\;X^\alpha \,\left( Y^\nu \,\eta ^{\beta \mu }+
Y^\mu \,\eta ^{\beta \nu }\right) +Y^\alpha \,\left( X^\nu
\,\eta ^{\beta \mu }+X^\mu \,\eta ^{\beta \nu }\right) $ \\
$T^{\mu\nu ,\, \alpha\beta}_7(X,Y)=Y^\nu \,X^\alpha \,X^\beta \,X^\mu +
Y^\mu \,X^\alpha \,X^\beta \,X^\nu +Y^\beta \,X^\alpha \,X^\mu \,X^\nu +
Y^\alpha \,X^\beta \,X^\mu \,X^\nu $ \\
$T^{\mu\nu ,\, \alpha\beta}_8(X,Y)=Y^\beta \,Y^\nu \,\eta ^{\alpha \mu }+
Y^\beta \,Y^\mu \,\eta ^{\alpha \nu }+
Y^\alpha \,Y^\nu \,\eta ^{\beta \mu }+
Y^\alpha \,Y^\mu \,\eta ^{\beta \nu }$ \\
$T^{\mu\nu ,\, \alpha\beta}_9(X,Y)=Y^\mu \,Y^\nu \,X^\alpha \,X^\beta +
Y^\alpha \,Y^\beta \,X^\mu \,X^\nu $ \\
$T^{\mu\nu ,\, \alpha\beta}_{10}(X,Y)=\left( Y^\beta \,X^\alpha +
Y^\alpha \,X^\beta \right) \,\left( Y^\nu \,X^\mu +Y^\mu \,X^\nu \right) $ \\
$T^{\mu\nu ,\, \alpha\beta}_{11}(X,Y)=Y^\beta \,Y^\mu \,Y^\nu \,X^\alpha +
Y^\alpha \,Y^\mu \,Y^\nu \,X^\beta +Y^\alpha \,Y^\beta \,Y^\nu \,X^\mu +
Y^\alpha \,Y^\beta \,Y^\mu \,X^\nu $ \\
$T^{\mu\nu ,\, \alpha\beta}_{12}(X,Y)=Y^\alpha \,Y^\beta \,Y^\mu \,Y^\nu $ \\
$T^{\mu\nu ,\, \alpha\beta}_{13}(X,Y)=Y^\mu \,Y^\nu \,\eta ^{\alpha \beta }+
Y^\alpha \,Y^\beta \,\eta ^{\mu \nu }$ \\
$T^{\mu\nu ,\, \alpha\beta}_{14}(X,Y)=\left( Y^\nu \,X^\mu +Y^\mu \,X^\nu
\right) \,\eta^{\alpha \beta }+\left( Y^\beta \,X^\alpha +
Y^\alpha \,X^\beta \right) \,\eta^{\mu \nu }$
   \end{tabular}
\vskip 1cm
\caption{A basis of 14 independent tensors
$T_i^{\mu\nu ,\alpha\beta}\left(X,Y\right)$
\label{table1}}
\end{table}

\begin{figure}
\caption{Lowest order matter contributions to the thermal 1PI graviton
two-point function. Curly lines denote the external gravitational field and
solid lines represent the  scalar particle.}\label{fig1}
\end{figure}

\begin{figure}
\caption{The forward scattering graphs corresponding to Fig. 1. Crossed graphs
with ($k\leftrightarrow -k$) are to be understood.}\label{fig2}
\end{figure}

\begin{figure}
\caption{One-loop contributions of radiation fields to the graviton
polarization tensor. Wavy lines denote the gauge field and broken
lines represent ghost particles.}\label{fig3}
\end{figure}

\begin{figure}
\caption{Forward scattering diagrams containing ghost particles connected
with Fig. 3. Crossed diagrams ($k\leftrightarrow -k$) should be included.}
\label{fig4}
\end{figure}

\begin{figure}
\caption{Lowest order contributions of graviton 1-point functions to
$\Pi_{tad}$. The black dot represents terms proportional to $\eta$.}
\label{fig5}
\end{figure}

\end{document}